\documentclass[fleqn,usenatbib]{mnras}

\usepackage{newtxtext,newtxmath}
\usepackage{subcaption}
\usepackage[T1]{fontenc}

\DeclareRobustCommand{\VAN}[3]{#2}
\let\VANthebibliography\thebibliography
\def\thebibliography{\DeclareRobustCommand{\VAN}[3]{##3}\VANthebibliography}

\usepackage{graphicx}	% Including figure files
\usepackage{amsmath}	% Advanced maths commands
\title[Ellipsoidal BH/NS candidates from \textit{Gaia} DR3]{Spectroscopic follow-up of black hole and neutron star candidates in ellipsoidal variables from \textit{Gaia} DR3}

\author[Nagarajan et al.]{
Pranav Nagarajan,$^{1}$\thanks{E-mail: pnagaraj@caltech.edu}
Kareem El-Badry,$^{1, 2}$
Antonio C. Rodriguez,$^{1}$
Jan van Roestel,$^{3, 1}$ and
Benjamin Roulston$^{1}$
\\
$^{1}$Department of Astronomy, California Institute of Technology, 1200 E. California Blvd., Pasadena, CA 91125, USA\\
$^{2}$Center for Astrophysics $|$ Harvard \& Smithsonian, 60 Garden Street, Cambridge, MA 02138, USA\\
$^{3}$Anton Pannekoek Institute for Astronomy, University of Amsterdam, 1090 GE Amsterdam, The Netherlands\\
}

\date{Accepted 2023 July 9. Received 2023 June 23; in original form 2023 April 12}

\pubyear{2023}

\begin{document}
\label{firstpage}
\pagerange{\pageref{firstpage}--\pageref{lastpage}}
\maketitle

\begin{abstract}
We present multi-epoch spectroscopic follow-up of a sample of ellipsoidal variables selected from {\it Gaia} DR3 as candidates for hosting quiescent black holes (BHs) and neutron stars (NSs). Our targets were identified as BH/NS candidates because their optical light curves -- when interpreted with models that attribute variability to tidal distortion of a star by a companion that contributes negligible light -- suggest that the companions are compact objects. From the likely BH/NS candidates identified in recent work accompanying \textit{Gaia} DR3, we select 14 of the most promising targets for follow-up. We obtained spectra for each object at 2-10 epochs, strategically observing near conjunction to best-constrain the radial velocity semi-amplitude. From the measured semi-amplitudes of the radial velocity curves, we derive minimum companion masses of $M_{2,\,\rm min} \leq 0.5 ~ M_{\odot}$ in all cases. Assuming random inclinations, the typical inferred companion mass is $M_2 \sim 0.15\,M_{\odot}$. This makes it unlikely that any of these systems contain a BH or NS, and we consider alternative explanations for the observed variability. We can best reproduce the observed light curves and radial velocities with models for unequal-mass contact binaries with starspots. Some of the objects in our sample may also be detached main-sequence binaries, or even single stars with pulsations or starspot variability masquerading as ellipsoidal variation. We provide recommendations for future spectroscopic efforts to further characterize this sample and more generally to search for compact object companions in close binaries. 
\end{abstract}


%%%%%%%%%%%%%%%%%%%%%%%%%%%%%%%%%%%%%%%%%%%%%%%%%%

%%%%%%%%%%%%%%%%% BODY OF PAPER %%%%%%%%%%%%%%%%%%

\section{Introduction}

 \label{sec:intro}

The Milky Way is thought to contain $\sim10^8$ stellar-mass black holes (BHs) and $\sim10^9$ neutron stars (NSs), an uncertain fraction of which are in binaries \citep[][]{Brown1994, Chawla2022}. Many theoretical models predict that binaries containing non-accreting BHs and NSs should significantly outnumber accreting systems, which are likely a rare outcome of binary evolution \citep[e.g.][]{PortegiesZwart1997}. However, almost all known stellar-mass BHs are accreting from a companion and were discovered via their emission at X-ray and/or radio wavelengths. Similarly, all unambiguously identified neutron stars are either in accreting binaries or are radio pulsars. There are several ongoing attempts to find dormant BHs and NSs in wider binaries, but it has become clear in recent years that these systems are relatively rare and false-positives are abundant \citep[e.g.][]{el_badry_2021, el_badry_burdge_2022, el_badry_2022a, el_badry_2022b}. To date, only $\sim$20 BHs have been dynamically confirmed \citep{McClintock2006, Corral-Santana2016}.

The recent 3rd data release of the {\it Gaia} mission \citep[``{\it Gaia} DR3'';][]{prusti_2016, vallenari_2022} is a treasure trove for studies of binary evolution and searches for rare objects. {\it Gaia} produces catalogs of astrometric, spectroscopic, and photometrically variable binaries \citep{eyer_2022, mowlavi_2022, halbwachs_2022, arenou_2022}. Several recent studies have searched for NS/BH companions in the astrometric and spectroscopic binary samples published in DR3 \citep[e.g.][]{shahaf_2023, gaia_bh1, gaia_bh2, chakrabarti_2023, badry_rix_2022, fu_2022}. In this study, we focus on the sample of ellipsoidal variable binaries detected via {\it Gaia} light curves. Ellipsoidal variability is sensitive to BH/NS companions in {\it close} binaries; i.e., the immediate progenitors of X-ray binaries.

Recently, \citet{gomel_faigler_mazeh_2021} introduced a new method to identify ellipsoidal variables hosting massive, unseen companions based on their light curves alone. Subsequently, \citet{gomel_gaia_2022} applied the method to the sample of candidate ellipsoidal variables published in DR3. They identified 262 sources whose light curves suggested the presence of an unseen companion more massive than the luminous primary, potentially indicative of a compact object. There has thus far been no spectroscopic follow-up of this sample of candidate BH/NS companions. While comparably large samples of ellipsoidal variables have been produced previously from photometric surveys before {\it Gaia} \citep[e.g. with OGLE; see][]{Gomel2021}, the {\it Gaia} DR3 sample is all-sky and dominated by relatively bright stars amenable for spectroscopic follow-up.

Since compact object binaries are an intrinsically rare population, vetting through spectroscopic follow-up is essential. This paper presents a spectroscopic pilot study of the \citet{gomel_gaia_2022} {\it Gaia} DR3 sample, with a goal of measuring dynamical masses and searching for luminous companions in a subset of their most promising candidates. This allows us to assess the purity of the full sample and diagnose the most common false-positives. 

The remainder of this paper is organized as follows. In Section \ref{sec:ellipsoidal}, we summarize the origin of ellipsoidal variability and the search strategy for high mass ratio companions proposed by \citet{gomel_faigler_mazeh_2021}. In Section \ref{sec:data}, we discuss how we selected the most promising variables for spectroscopic follow-up and describe the follow-up observations. In Section \ref{sec:analysis}, we analyze the radial velocities and infer companion masses. Finally, in Section \ref{sec:discussion}, we discuss possible failure modes of the selection strategy used by \citet{gomel_gaia_2022} including contact binaries, binaries containing spotted stars, and single stars with pulsational or starspot variability mistaken for tidal deformation. We summarize our results in Section \ref{sec:conclusion}. 

\section{A Minimum Mass Ratio Inferred From Ellipsoidal Variability}
\label{sec:ellipsoidal}

In a close binary, tidal forces can distort one or both stars into a teardrop shape. The resulting time-variable geometric cross-section leads to photometric variability with a period of half the orbital period. The amplitude of this ``ellipsoidal'' variability depends on the orbital inclination, the mass ratio $q = M_2 / M_1$, where $M_1$ is the star dominating the optical light, and the Roche lobe filling factor of the distorted luminous primary \citep[$M_1$; e.g.][]{morris_naftilan_1993}. In general, the variability amplitude is largest for edge-on inclinations and is largest in binaries with large mass ratios and Roche lobe filling factors approaching unity.

While the inclination and Roche lobe filling factor are generally unknown, they cannot be larger than 90 degrees and unity, respectively. Recently, \citet{gomel_faigler_mazeh_2021} demonstrated that, under the assumption that all variability is indeed due to tidal deformation, this fact can be used to infer a stringent lower limit on the mass ratio of an ellipsoidal variable. This quantity, which they term the ``modified minimum mass ratio'' (mMMR or $\hat{q}_{\min}$), is robust to assumptions about the mass, radius, and evolutionary state of the component stars \citep{gomel_faigler_mazeh_2020, gomel_faigler_mazeh_2021}.  In brief, Gomel et al. show that under the assumptions of their model, a variability half-amplitude larger than $\gtrsim$0.1 mag can only occur if the mass ratio is larger than one; i.e., if the unseen companion is more massive than the source itself. This reflects the fact that a Roche-lobe filling star is more tidally distorted when the mass ratio is large than when it is small. The expected variability amplitude in a given photometric band can be calculated straightforwardly for a given mass ratio, inclination, and Roche lobe filling factor using tabulated approximations \citep[e.g.][]{morris_naftilan_1993, gomel_faigler_mazeh_2020}.

\citet{gomel_gaia_2022} analyzed the {\it Gaia} DR3 photometrically-variable binaries using this strategy to identify binaries with large $\hat{q}_{\min}$. Starting with $\sim20$ million possible ellipsoidal systems, and only including systems with well-constrained periods $< 2.5$ days, reliable and significant variability amplitudes, and $\hat{q}_{\min} > 0.5$, they identified 262 high-confidence ellipsoidal variables near the main-sequence whose variability amplitudes imply mass ratios larger than unity at the one-sigma level. Their sample is restricted to sources near the main-sequence in the color-magnitude diagram, with the goal of excluding mass-transfer binaries in which $q > 1$, but the companion is a main-sequence star \citep[``Algol-type'' binaries; e.g.][]{Clavel2021, el_badry_2022b}.

\section{Data} \label{sec:data}

\subsection{Sample Selection}

From the 6306 probable ellipsoidal variables identified by \citet{gomel_gaia_2022}, we selected 14 bright candidates with high minimum mass ratios for spectroscopic follow-up. We limited the sample to variables with apparent $G$-band magnitude $<$ 17 and well-constrained distances (\texttt{parallax\_over\_error} $>$ 3). We further restricted the sample to targets with declination $\delta >-30$ deg, which were observed by the Pan-STARRS survey \citep{chambers_2016} and thus have well-constrained extinctions from the 3D dust map of \citet{green_2019}. Then, we investigated the variables with $\hat{q}_{\min} \geq$ 0.8 by hand, verifying that their optical light curves from Public Data Release 12 of the Zwicky Transient Facility \citep[ZTF;][]{bellm_2019} (queried from IRSA using \texttt{ztfquery}; see \citealt[][]{rigault_ztfquery}) indeed appear to be dominated by ellipsoidal variability.

\begin{figure*}
\includegraphics[width=\textwidth]{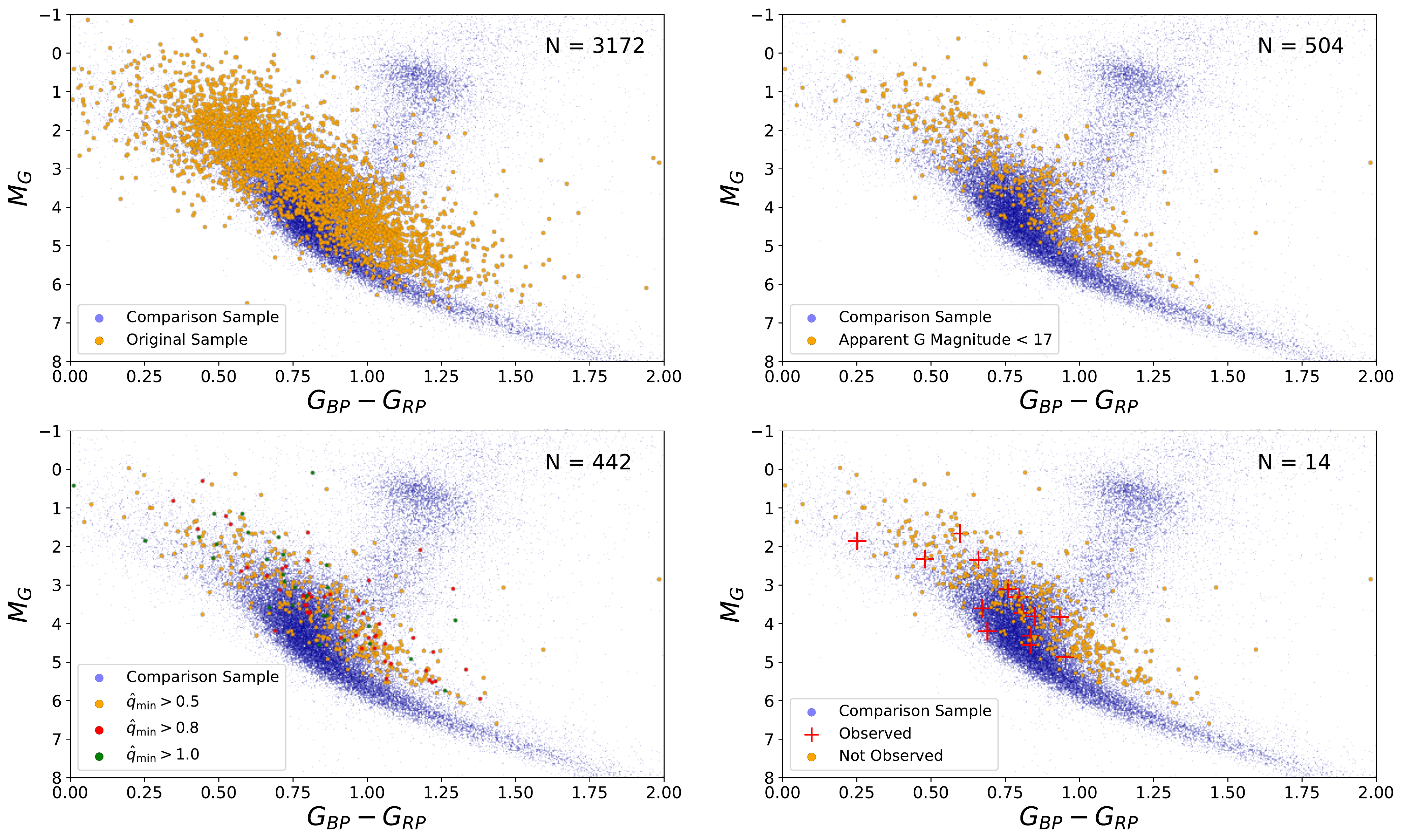}
\caption{Color-magnitude diagrams visualizing our sample selection. The upper left panel displays the original sample of ellipsoidal variables (i.e.\ objects in the \citet{gomel_gaia_2022} catalog that have available extinctions) plotted over a random comparison sample of $\sim45000$ stars in {\it Gaia} DR3 with declination $\delta > -30^{\circ}$ and apparent \textit{Gaia} $G$-band magnitude $G < 17$. The upper right panel displays the sample of ellipsoidal variables with apparent magnitude $G < 17$. The lower left panel displays all the ellipsoidal variables in our vetted sample (i.e., objects with well-constrained distances and well-sampled ZTF light curves) colored by minimum mass ratio. The lower right panel shows our final set of observed targets (red) and objects that were either rejected after inspection of their light curves or not observed (yellow). Note that a significant fraction of all candidates are above the main-sequence, as expected for luminous binaries. Our follow-up prioritized objects near the main-sequence, which are more likely to host dark companions.}
\label{fig:sample}
\end{figure*}

It was immediately clear that a significant fraction of the candidates fall above the main sequence in the extinction-corrected color-magnitude diagram (Figure \ref{fig:sample}), as is expected for binaries containing two luminous stars that both contribute significantly to the total flux. Such binaries are unlikely to host massive dark companions, so we prioritized sources that are not significantly above the main sequence in our spectroscopic follow-up. 

The series of cuts leading to our final sample of 14 probable ellipsoidal variables with spectroscopic observations is displayed on the de-reddened {\it Gaia} CMD in Figure \ref{fig:sample}, with the blue points in the background of each CMD representing a random comparison sample of $\sim45000$ stars for {\it Gaia} DR3 with declination $\delta > -30^{\circ}$ and apparent \textit{Gaia} $G$-band magnitude $G < 17$. The initial cut (which restricted the sample to bright variables with $G < 17$) was most important, as it reduced the size of the initial sample of variables with available extinctions by more than a factor of 6. While most of the candidates in this sample also had well-constrained distances (i.e. a relative parallax / error $> 3$) and well-sampled ZTF light curves (i.e.\ $\gtrsim 20$ epochs in the ZTF $g$-band), we found that a majority of the sample still lied above the main sequence. Hence, from this vetted sample, we chose a set of 12 promising variables near the main sequence of the comparison sample that had been identified by \citet{gomel_gaia_2022} to have inferred $\hat{q}_{\min} > 0.8$. For the sake of completeness, and to investigate the effect of selection bias, we also observed two candidates with $0.5 < \hat{q}_{\min} < 0.8$. Our final sample of 14 candidate ellipsoidal variables has apparent $G$-band magnitudes between 16 and 17 and orbital periods ranging from about $0.27$ days to about $0.75$ days. They are located on the upper main sequence on the CMD (i.e.\ bluer than $G_{BP} - G_{RP} = 1$) and have likely primary masses between $1-2 ~ M_{\odot}$ (see Section \ref{ssec:companions} for details). Their light curve-inferred minimum mass ratios are typically $\approx 1.0$, as might be expected for NS or low-mass BH companions.

\begin{figure*}
\includegraphics[width=\textwidth]{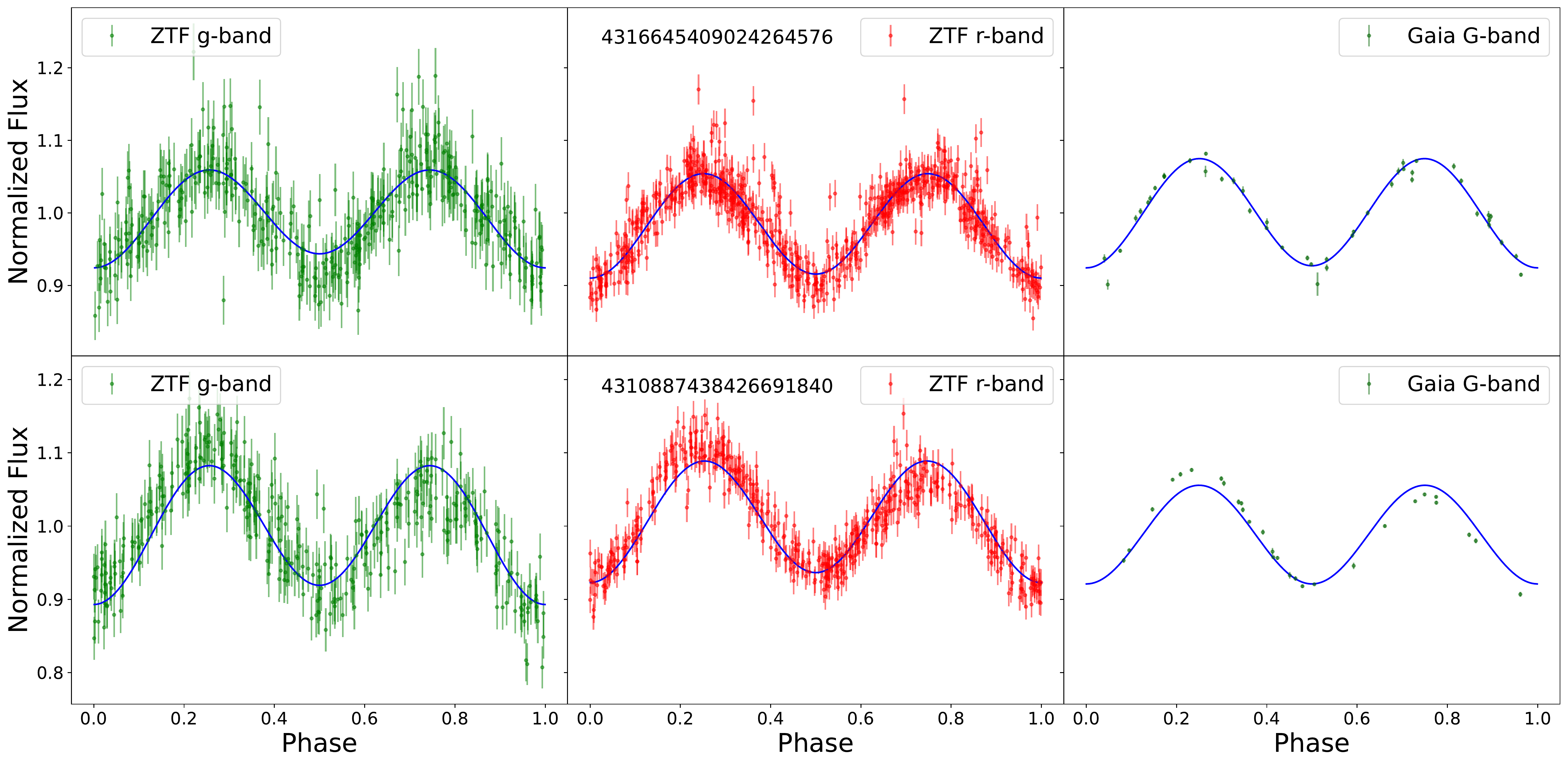}
\caption{Comparison of light curve data in ZTF $g$-band, ZTF $r$-band, and \textit{Gaia} $G$-band for ellipsoidal variables. Top Row: Light curves for typical ellipsoidal variable candidate with \textit{Gaia} DR3 ID 4316645409024264576. Bottom Row: Light curves for anomalous ellipsoidal variable candidate with \textit{Gaia} DR3 ID 4310887438426691840, rejected due to significant difference between light curve maxima (a signature of the ``O'Connell Effect'', e.g.\ see \citealt[][]{oconnell_1951} and \citealt[][]{pribulla_2011}). A summary of our reasons for the rejection of candidates with $\hat{q}_{\min} > 0.8$ which otherwise satisfy our automated selection criteria can be found in Appendix \ref{appendix:rejection}.  Details on our best-fit light curves (which model ellipsoidal variability as the sum of Fourier harmonics) can be found in Appendix \ref{appendix:lcs}. For these targets and many other in our sample, the ZTF light curves are better-sampled than the relatively sparse {\it Gaia} light curves, but the {\it Gaia} light curves have higher SNR.}
\label{fig:lc_comp}
\end{figure*}

As an example, we provide ZTF $g$-band, ZTF $r$-band, and \textit{Gaia} $G$-band light curves for a typical ellipsoidal variable candidate and an anomalous ellipsoidal variable candidate in Figure \ref{fig:lc_comp}. For the typical ellipsoidal variable candidate, the characteristic periodicity and difference in light curve minima is evident; for this variable, the peak-to-peak variability amplitude is 0.135, and the corresponding modified minimum mass ratio (mMMR or $\hat{q}_{\min}$) inferred by \citet{gomel_gaia_2022} is $1.01 \pm 0.26$. On the other hand, the anomalous ellipsoidal variable candidate displays an uncharacteristic difference in light curve maxima, and was thus not considered in our final sample. In both cases, it is clear that using ZTF photometry often represents an improvement over the relatively sparsely sampled {\it Gaia} light curves. On the other hand, the photometric precision of the {\it Gaia} light curves is generally higher than that of the ZTF data.

\subsection{Follow-up Strategy}

We obtained follow-up spectra at multiple epochs using spectrographs at Palomar and Keck Observatories, with the goal of measuring multi-epoch radial velocities (RVs) with a target precision of $\approx 10\,\rm km\,s^{-1}$ at $G = 17$. Since the expected epoch-to-epoch RV shifts for a close BH or NS companion are large ($\gtrsim 100\,\rm km\,s^{-1}$), this modest RV precision is sufficient to test the hypothesis that the targets have a BH or NS companion.

We mainly relied on the Double Spectrograph \citep[DBSP;][]{Oke_1982} on the Palomar 200-inch telescope. Observing on the red and blue arms simultaneously, we used a $1.0, 1.5$, or $2.0$ arcsec slit (depending on seeing conditions) to obtain spectra over a total wavelength range of 3500 $\text{\AA}$ to 8000 $\text{\AA}$ with a typical spectral resolution $R\approx 2000$. Data reduction (i.e.\ bias and flat field correction, cosmic ray removal, wavelength calibration, sky subtraction, extraction of 1D spectra, and heliocentric RV corrections) was performed using PypeIt \citep{Prochaska_2020}. 

We also observed some targets using the Low Resolution Imaging Spectrometer \citep[LRIS;][]{Oke_1995} and the Echellete Spectrograph and Imager \citep[ESI;][]{Sheinis_2002} at Keck.
For these instruments, data reduction was performed using the LRIS automated reduction Pipeline \citep[LPIPE;][]{perley_2019} and the MAuna Kea Echelle Extraction (MAKEE) pipeline, respectively. A log of all of our observations can be found in Table \ref{tab:epoch_rvs}.

\subsection{Flexure Corrections}

The DBSP instrument hangs off of the back of the Hale telescope, and can thus move from epoch to epoch. This ``flexure'' can lead to changes in the wavelength solution between calibration arcs and science exposures, ultimately resulting in spurious radial velocities. A common strategy to minimize flexure is to take an arc lamp exposure on-sky immediately before or after observing each target. However, this approach significantly increases the required telescope time, and we found that even when arcs were taken immediately after each target, the inferred RVs of standard stars were only stable at the $\approx 30\,\rm km\,s^{-1}$ level. We therefore used telluric absorption lines with known wavelengths as a reference instead. Specifically, we produced a telluric model spectrum with appropriate airmass for each observation from a grid of high-resolution models of telluric absorption spectra from the HITRAN2020 molecular spectroscopic database \citep{gordon_2022}. 

\begin{figure*}
\includegraphics[width=\textwidth]{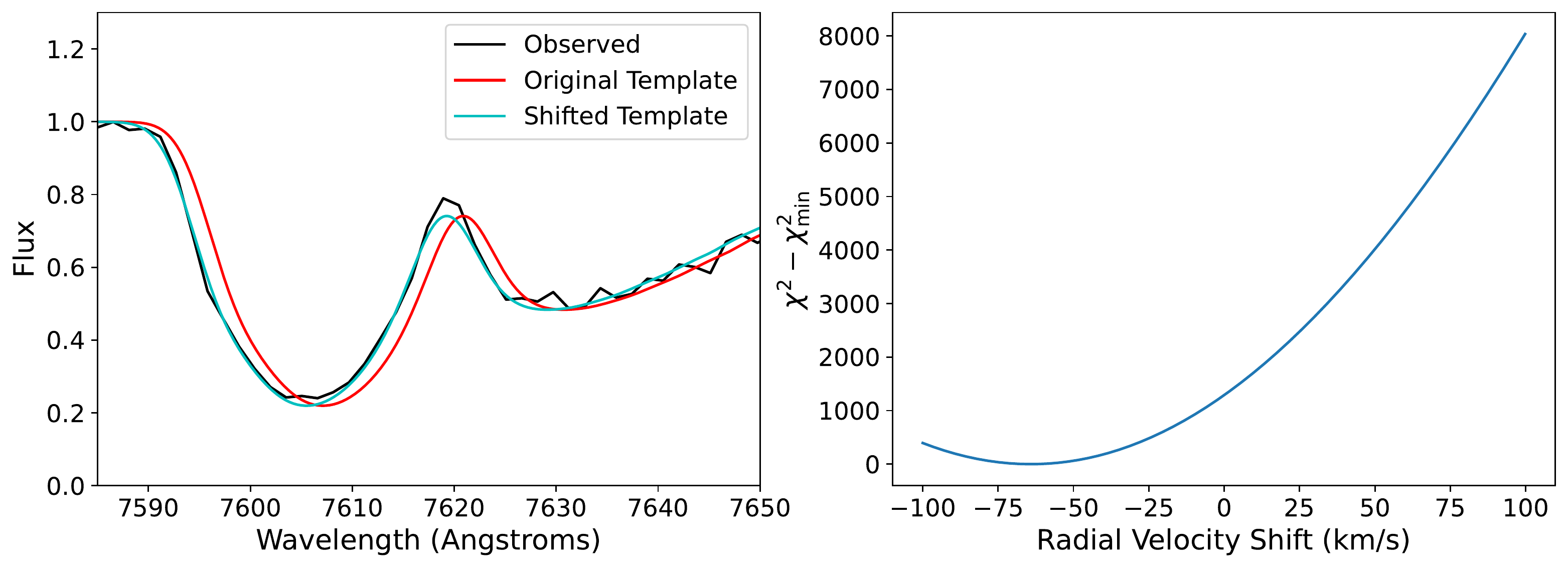}
\caption{Determination of flexure correction for DBSP observation of ellipsoidal variable candidate with \textit{Gaia} DR3 ID 2062434366319612672 at UTC epoch 2022-08-25 07:48:34. The left panel shows the effect of applying the determined flexure correction to the template spectrum; we see that the telluric line in the shifted template better matches the telluric line in the observed spectrum. The right panel shows the chi-squared value as a function of radial velocity shift. The minimum of the chi-squared curve was identified as the inferred flexure correction, which is -64 km s$^{-1}$ at this epoch. Flexure corrections significantly improved the stability of our RV measurements.}
\label{fig:flexure}
\end{figure*}

We smoothed this template to the resolution of the observation assuming a Gaussian line spread function, and determined the radial velocity shift of the telluric template that minimized the $\chi^2$ statistic with respect to the observed spectrum (see Equation \ref{eq:chi2}) within a wavelength range of 7585 $\text{\AA}$ to 7650 $\text{\AA}$ (which contains the A-band of molecular oxygen and is commonly used for telluric wavelength calibrations; e.g. \citealt{Chubak2012}). This allowed us to calculate flexure corrections as a pre-processing step before applying heliocentric corrections and measuring RVs at each epoch.  Once our flexure corrections were applied, the inferred RVs of standard stars were generally consistent within 10 km/s. We present an example of our flexure correction method in Figure \ref{fig:flexure}. Our flexure correction method was only necessary for DBSP; we found that the ESI RVs were stable, and that the flexure correction built into LPipe (which uses sky emission lines) worked well for the LRIS observations.

\section{Analysis}
\label{sec:analysis}

\subsection{Radial Velocity Measurements}

We used theoretical grids provided by \citet{pecaut_2013} to estimate the spectral type and effective temperature of each primary star based on its $G_{BP} - G_{RP}$ color. We retrieved the appropriate BOSZ Kurucz model spectrum for each target (with appropriate instrumental broadening) from the STSCI Archive \citep{bohlin_2017}. Since our candidates are all rapidly rotating, there are not many strong metal lines in their observed spectra. Hence, we assumed a solar metallicity for all model spectra; nevertheless, our results are not strongly sensitive to this choice. Adopting a linear limb darkening coefficient of 0.5, we then used the \texttt{rotBroad} function from PyAstronomy \citep{pya}, which implements the formulae in  \citet{gray_1992}, to apply rotational broadening to the model spectra assuming a projected rotation velocity $v \sin{i} = 2 \pi R / P$. Here, the radius was assumed to follow a typical main sequence mass-radius relation given by $(M / M_{\odot}) = (R / R_{\odot})^{0.8}$ for $M < 1 ~ M_{\odot}$ and $(M / M_{\odot}) = (R / R_{\odot})^{0.57}$ for $M \geq 1 ~ M_{\odot}$, and the mass was inferred from the absolute $G-$band magnitude (see Section \ref{ssec:companions}). We found that this rotational broadening was typically on the order of $v \sin{i} \sim 100$ km s$^{-1}$. The result of this procedure was a template spectrum with absorption lines that had a similar depth and width to the observed spectra.

To determine the best fit radial velocity, we found the minimum of the $\chi^2$ statistic between the shifted template spectrum and observation as a function of radial velocity. The chi-squared statistic is defined to be:

\begin{equation}
\label{eq:chi2}
\chi^2 = \sum_{\lambda} \frac{(f_{\text{template}, \lambda} - f_{\text{obs}, \lambda})^2}{\sigma_\lambda^2}
\end{equation}

where $f_{\text{template}, \lambda}$ is the flux of the template spectrum, $f_{\text{obs}, \lambda}$ is the flux of the observed spectrum, and $\sigma_\lambda$ is the uncertainty in the flux of the observed spectrum at wavelength $\lambda$, respectively. Note that flexure corrections were subtracted from our determined radial velocity to achieve the final RV measurement at a particular epoch. We repeated this procedure for each epoch for each ellipsoidal variable.

\begin{figure*}
\includegraphics[width=\textwidth]{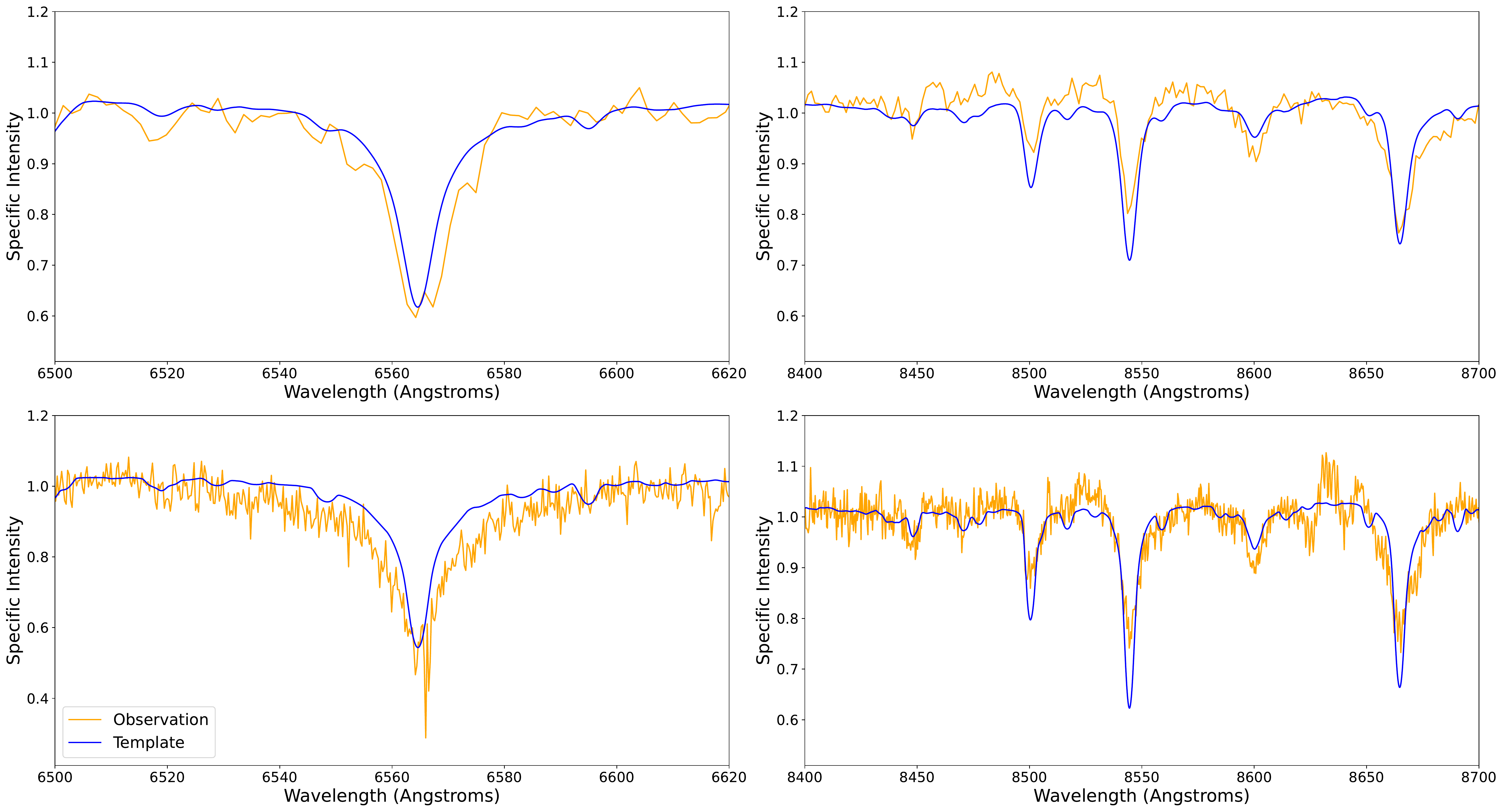}
\caption{Example spectral fits for a DBSP observation and an ESI observation of ellipsoidal variable with \textit{Gaia} DR3 ID 2062434366319612672. In both cases, we choose wavelength ranges for template fitting to emphasize the prominent signals due to the H$\alpha$ line and the Ca triplet. The agreement between the data and the model is not perfect, but is sufficient for robust measurement of radial velocities.}
\label{fig:typical_spectrum}
\end{figure*}

We provide example spectral fits for a DBSP (i.e.\ low-resolution) observation and an ESI (higher-resolution) observation of a candidate ellipsoidal variable in Figure \ref{fig:typical_spectrum}. We compare the two spectral fits over wavelength ranges which include either the H$\alpha$ line (i.e.\ 6000 $\text{\AA}$ - 6600 $\text{\AA}$) or the Ca triplet (i.e.\ 8400 $\text{\AA}$ - 8900 $\text{\AA}$). After testing both the stability of our RV determinations and the relative strength of the signals, we decided to rely on the H$\alpha$ line and Ca triplet for RV determination in DBSP/LRIS and ESI observations, respectively. In most cases, we found the the template spectra to match the shape of the observed spectra reasonably well. In a few cases, we found via visual inspection that the spectral model did not yield a reliable fit in the default spectral region (i.e.\ H$\alpha$ or Ca triplet) due to data artifacts and/or an imperfect choice of template. In these cases, we used the other spectral region instead, regardless of instrument. A few observed spectra showed hints of line splitting and/or epoch-to-epoch changes in line profile, which are most likely due to a luminous secondary. We did not attempt to account for these variations when inferring RVs. We provide a summary of all of our RV measurements in Table \ref{tab:epoch_rvs}. The reported RV uncertainties are based on observations of standard stars with the same instrumental setup; in all cases, we adopt an uncertainty of $10\,\rm km\,s^{-1}$ for DBSP and LRIS observations, and $5\,\rm km\,s^{-1}$ for ESI observations.

\begin{table*}
\caption{Summary of radial velocity measurements for vetted sample of ellipsoidal variables.} \label{tab:epoch_rvs}
\begin{tabular}{cccc}
\hline
\textit{Gaia} Source ID & Observation Time (HJD) & Radial Velocity (km s$^{-1}$) & Instrument \\
\hline
2033460169042378752 & 2459818.8998 & 26$\pm$10 & LRIS \\
2033460169042378752 & 2460084.8948 & 8$\pm$10 & DBSP \\
2033460169042378752 & 2460055.9840 & 57$\pm$10 & DBSP \\
2033460169042378752 & 2459824.8937 & 37$\pm$5 & ESI \\
2033460169042378752 & 2459818.9928 & 55$\pm$10 & LRIS \\
1342190925113500416 & 2459980.0472 & -16$\pm$10 & DBSP \\
1342190925113500416 & 2460055.9698 & -23$\pm$10 & DBSP \\
1342190925113500416 & 2459967.0470 & -65$\pm$10 & DBSP \\
3396537760623971328 & 2459878.1285 & 12$\pm$5 & ESI \\
3396537760623971328 & 2459877.9158 & 12$\pm$5 & ESI \\
3396537760623971328 & 2459878.0306 & 22$\pm$5 & ESI \\
4277589209792837632 & 2459813.7896 & -97$\pm$10 & DBSP \\
4277589209792837632 & 2460055.8814 & -105$\pm$10 & DBSP \\
4277589209792837632 & 2460084.8819 & -106$\pm$10 & DBSP \\
4277589209792837632 & 2459813.6924 & -93$\pm$10 & DBSP \\
2057298203894617088 & 2459813.8107 & -25$\pm$10 & DBSP \\
2057298203894617088 & 2459877.7123 & -58$\pm$5 & ESI \\
2057298203894617088 & 2459851.6926 & 6$\pm$10 & DBSP \\
2057298203894617088 & 2459813.9245 & -43$\pm$10 & DBSP \\
2211471926201876992 & 2459877.7954 & -93$\pm$5 & ESI \\
2211471926201876992 & 2459877.9057 & -113$\pm$5 & ESI \\
4316645409024264704 & 2460084.9442 & -54$\pm$10 & DBSP \\
4316645409024264704 & 2459873.7367 & 2$\pm$10 & DBSP \\
4316645409024264704 & 2459813.8025 & -5$\pm$10 & DBSP \\
4316645409024264704 & 2459813.7065 & -3$\pm$10 & DBSP \\
188218592331364736 & 2459970.6299 & -13$\pm$10 & DBSP \\
188218592331364736 & 2459969.6031 & 22$\pm$10 & DBSP \\
188218592331364736 & 2459878.1472 & 12$\pm$5 & ESI \\
188218592331364736 & 2459878.0710 & 22$\pm$5 & ESI \\
188218592331364736 & 2459877.8848 & 22$\pm$5 & ESI \\
188218592331364736 & 2459851.8790 & 11$\pm$10 & DBSP \\
188218592331364736 & 2459847.1070 & 3$\pm$10 & LRIS \\
188218592331364736 & 2459819.1302 & 8$\pm$10 & LRIS \\
188218592331364736 & 2459818.1245 & 6$\pm$10 & LRIS \\
4305376170746099712 & 2460055.9377 & -47$\pm$10 & DBSP \\
4305376170746099712 & 2459816.9013 & -58$\pm$10 & DBSP \\
4305376170746099712 & 2459816.8212 & -28$\pm$10 & DBSP \\
3456522854428709888 & 2459971.8484 & 34$\pm$10 & DBSP \\
3456522854428709888 & 2459851.8872 & 15$\pm$10 & DBSP \\
3456522854428709888 & 2459818.1288 & 8$\pm$10 & LRIS \\
3456522854428709888 & 2459819.1384 & 26$\pm$10 & LRIS \\
3456522854428709888 & 2459878.0885 & 12$\pm$5 & ESI \\
3456522854428709888 & 2459877.9523 & 42$\pm$5 & ESI \\
3456522854428709888 & 2459874.0142 & 57$\pm$10 & DBSP \\
3457961423656077312 & 2459815.9623 & 0$\pm$10 & DBSP \\
3457961423656077312 & 2459847.1222 & 13$\pm$10 & LRIS \\
3457961423656077312 & 2459877.8933 & -2$\pm$5 & ESI \\
3457961423656077312 & 2459878.0804 & -58$\pm$5 & ESI \\
3457961423656077312 & 2459813.9804 & -52$\pm$10 & DBSP \\
2163873350591724032 & 2459817.8809 & -71$\pm$10 & LRIS \\
2163873350591724032 & 2459818.8784 & -63$\pm$10 & LRIS \\
2163873350591724032 & 2459824.8820 & 63$\pm$5 & ESI \\
2163873350591724032 & 2459877.7848 & 42$\pm$5 & ESI \\
2062434366319612672 & 2459816.8319 & 16$\pm$10 & DBSP \\
2062434366319612672 & 2459816.9133 & -62$\pm$10 & DBSP \\
2062434366319612672 & 2459877.7710 & -73$\pm$5 & ESI \\
2062434366319612672 & 2459873.7703 & 29$\pm$10 & DBSP \\
4310698008860121088 & 2459846.8681 & 18$\pm$10 & LRIS \\
4310698008860121088 & 2459877.7400 & 27$\pm$5 & ESI \\
4310698008860121088 & 2460084.9352 & 17$\pm$10 & DBSP \\
\hline
\end{tabular}
\end{table*}

\subsection{Radial Velocity Fits}

We fit for the semi-amplitude $K$ and center-of-mass radial velocity $\gamma$ based on a sinusoidal function:

\begin{equation}
    RV(t) = \gamma - K \sin{\left(\frac{2 \pi (t - T_0)}{P}\right)}
\end{equation}

This model assumes that the orbit is circular, which is justified, since ellipsoidal binaries undergo tidal interactions that both circularize the orbit and synchronize the rotational and orbital periods \citep{Zahn1977, Lurie_2017}. The orbital period $P$ is fixed to the value from the \textit{Gaia} DR3 catalog, while the epoch of conjunction $T_0$ is determined from the ZTF light curve in the $g$-band (see Appendix \ref{appendix:lcs} for details). Since there are two free parameters, we only analyzed variables with at least two RV measurements. 

In general, we expect that $K \geq 0$. If we instead inferred $K < 0$ from our initial fit, then we took this as an indication that our assumed $T_0$ was inaccurate by half a period (i.e.\ the deeper minimum of the ZTF $g$-band light curve does not actually correspond to the true epoch of conjunction). While this could physically occur due to, e.g., a large spot on the side of the luminous component that faces away from an unseen massive companion, it is more likely that the uncertainty in the value of $T_0$ is simply a consequence of the light curve minima having similar depths. In such cases, spectroscopic follow-up is necessary to provide further information and help identify the epoch of conjunction. Candidates for which we shifted $T_0$ by half a period are indicated using an asterisk in Appendix \ref{appendix:lcs}.

To fit this model to the observed RVs, we performed MCMC sampling with the \texttt{emcee} package \citep{Foreman-Mackey_2013}. We used uniform priors over a wide range of possible values, restricting $\lvert \gamma \rvert < 500$ and $0 < K < 1000$. The likelihood compares measured and predicted RVs assuming Gaussian uncertainties. We randomly initialized 100 walkers and sampled the posterior for 1000 iterations. We report the median and middle 68\% of the posterior samples for both quantities in Table \ref{tab:results}. In cases where we only have two measurements, the RV curves may not be well-constrained, and the error bars on our estimates of the companion mass reflect this uncertainty. Even in these cases, the measured RVs are -- when combined with the photometric ephemeris -- strongly inconsistent with a companion mass $\gtrsim 1\,M_{\odot}$, and thus make it unlikely that the companions are neutron stars or black holes.

\begin{table*}
\caption{Derived companion masses for vetted sample of ellipsoidal variables, assuming an inclination of 90$^{\circ}$. Note that $\hat{q}_{\min}$ is the modified minimum mass ratio reported by \citet{gomel_faigler_mazeh_2021}, and is not what we derived based on our radial velocity measurements.}
\label{tab:results}
\begin{tabular}{cccccccc}
\hline
\textit{Gaia} Source ID & Period (days) & $T_0$ (HJD) & $\gamma$ (km s$^{-1}$) & $K$ (km s$^{-1}$) & $M_1$ ($M_{\odot}$) & $M_{2, \min}$ ($M_{\odot}$) & $\hat{q}_{\min}$ \\
\hline
2033460169042378752 & $0.270854\pm0.000057$ & $2458206.1367\pm0.0012$ & $41^{+4}_{-4}$ & $22^{+6}_{-6}$ & $1.14\pm0.11$ & $0.07^{+0.02}_{-0.02}$ & $0.91\pm0.22$ \\
1342190925113500416 & $0.279545\pm0.000078$ & $2458204.9995\pm0.0008$ & $-31^{+6}_{-6}$ & $17^{+6}_{-7}$ & $0.97\pm0.09$ & $0.05^{+0.02}_{-0.02}$ & $0.52\pm0.12$ \\
3396537760623971328 & $0.311336\pm0.000074$ & $2458205.7569\pm0.0016$ & $16^{+3}_{-3}$ & $7^{+4}_{-4}$ & $1.11\pm0.11$ & $0.02^{+0.01}_{-0.01}$ & $0.69\pm0.17$ \\
4277589209792837632 & $0.329189\pm0.000082$ & $2458206.0749\pm0.0014$ & $-100^{+5}_{-5}$ & $8^{+5}_{-6}$ & $1.49\pm0.14$ & $0.04^{+0.02}_{-0.03}$ & $1.18\pm0.34$ \\
2057298203894617088 & $0.352784\pm0.000096$ & $2458207.0088\pm0.0011$ & $-5^{+8}_{-8}$ & $52^{+9}_{-10}$ & $1.25\pm0.12$ & $0.22^{+0.04}_{-0.05}$ & $1.01\pm0.26$ \\
2211471926201876992 & $0.353515\pm0.000108$ & $2458230.0924\pm0.0011$ & $-99^{+4}_{-4}$ & $14^{+5}_{-5}$ & $1.28\pm0.12$ & $0.06^{+0.02}_{-0.02}$ & $0.82\pm0.20$ \\
4316645409024264704 & $0.390803\pm0.000094$ & $2458207.0437\pm0.0018$ & $-10^{+5}_{-5}$ & $16^{+6}_{-6}$ & $1.77\pm0.78$ & $0.08^{+0.04}_{-0.05}$ & $1.01\pm0.25$ \\
188218592331364736 & $0.395750\pm0.000142$ & $2458206.6547\pm0.0011$ & $15^{+2}_{-2}$ & $5^{+3}_{-4}$ & $1.04\pm0.10$ & $0.02^{+0.01}_{-0.01}$ & $1.08\pm0.27$ \\
4305376170746099712 & $0.409171\pm0.000104$ & $2458279.0047\pm0.0027$ & $-39^{+6}_{-7}$ & $17^{+9}_{-9}$ & $1.25\pm0.12$ & $0.07^{+0.04}_{-0.04}$ & $1.07\pm0.29$ \\
3456522854428709888 & $0.430926\pm0.000134$ & $2458205.7704\pm0.0009$ & $32^{+3}_{-3}$ & $21^{+4}_{-4}$ & $1.32\pm0.13$ & $0.09^{+0.02}_{-0.02}$ & $1.04\pm0.27$ \\
3457961423656077312 & $0.458269\pm0.000152$ & $2458385.8186\pm0.0022$ & $-26^{+3}_{-3}$ & $37^{+4}_{-4}$ & $1.42\pm0.14$ & $0.18^{+0.02}_{-0.02}$ & $1.14\pm0.28$ \\
2163873350591724032 & $0.481034\pm0.000174$ & $2458303.9553\pm0.0032$ & $-8^{+4}_{-4}$ & $61^{+4}_{-4}$ & $2.06\pm0.08$ & $0.41^{+0.03}_{-0.03}$ & $1.04\pm0.26$ \\
2062434366319612672 & $0.721862\pm0.000404$ & $2458207.1999\pm0.0012$ & $-33^{+4}_{-4}$ & $44^{+4}_{-4}$ & $1.76\pm0.77$ & $0.30^{+0.09}_{-0.10}$ & $1.06\pm0.26$ \\
4310698008860121088 & $0.751384\pm0.000436$ & $2458205.9044\pm0.0043$ & $32^{+6}_{-7}$ & $17^{+11}_{-12}$ & $2.20\pm0.09$ & $0.12^{+0.08}_{-0.10}$ & $1.62\pm0.46$ \\
\hline
\end{tabular}
\end{table*}

\begin{figure*}
\centering
\begin{subfigure}{.49\textwidth}
  \centering
  \includegraphics[width=\columnwidth]{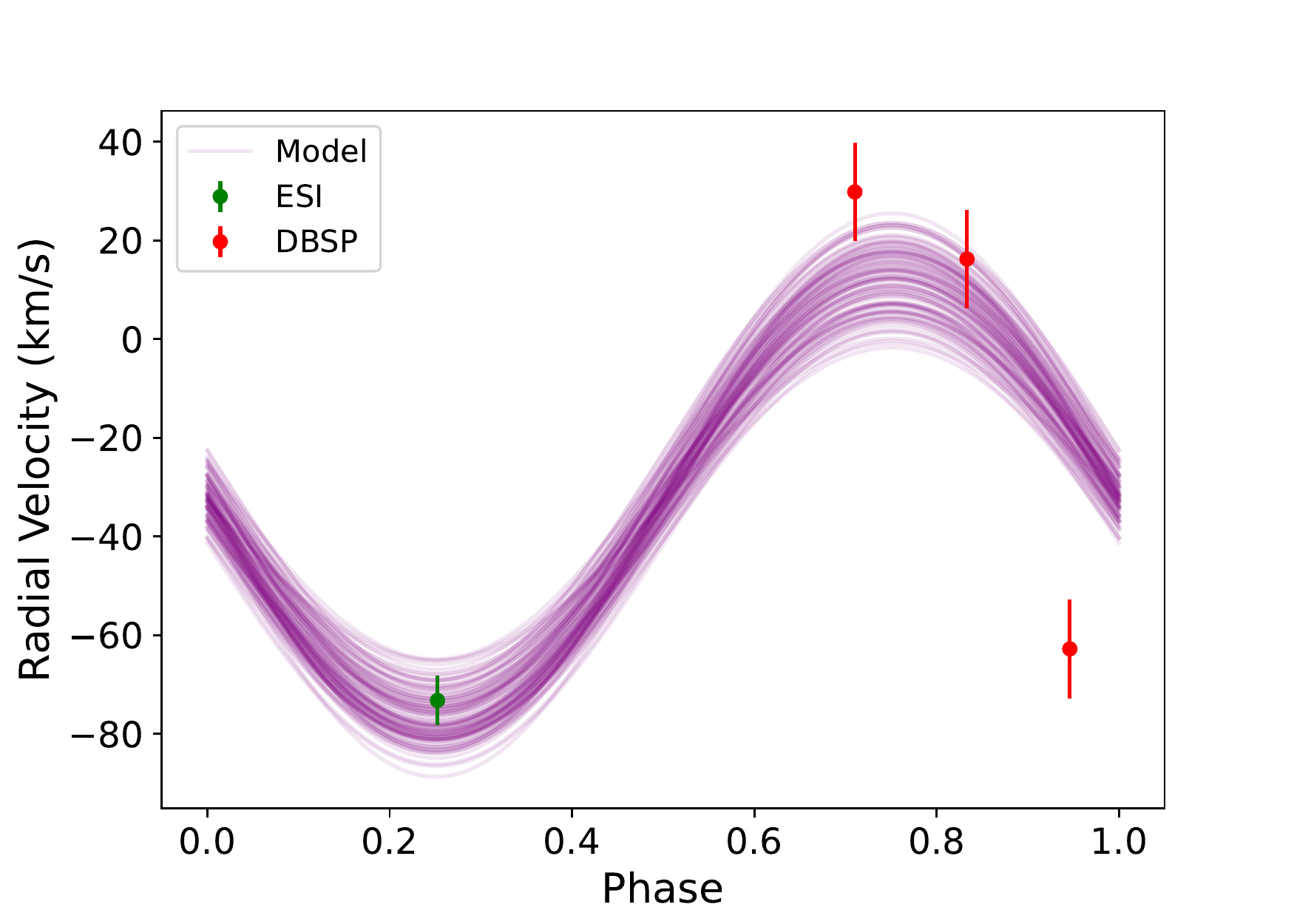}
\end{subfigure}
\begin{subfigure}{.49\textwidth}
  \centering
  \includegraphics[width=\columnwidth]{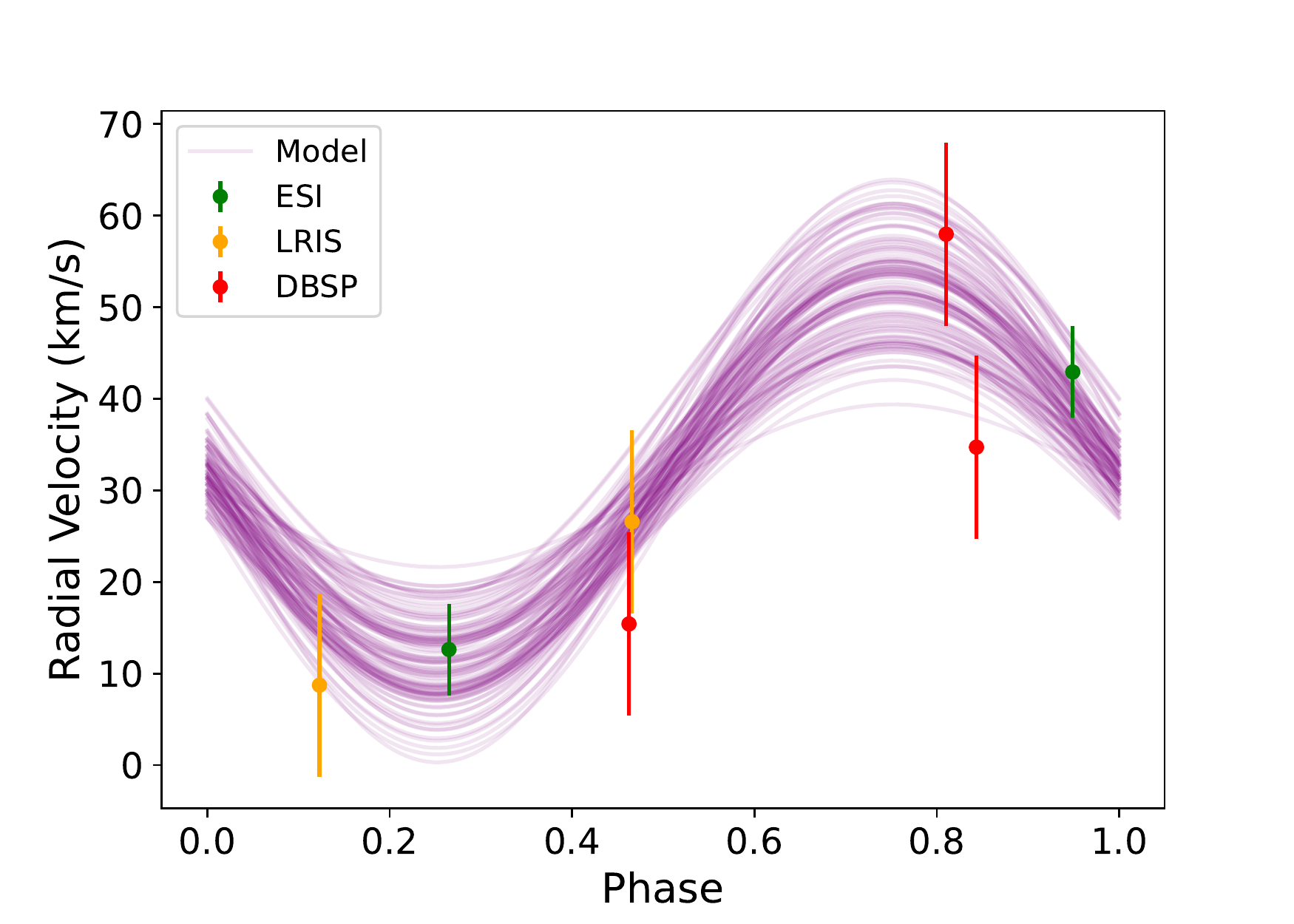}
\end{subfigure}
\caption{Left Panel: Radial velocity curve for ellipsoidal variable candidate with \textit{Gaia} DR3 ID 2062434366319612672. The inferred RV semi-amplitude is 44 km s$^{-1}$. Right Panel: Radial velocity curve for ellipsoidal variable candidate with \textit{Gaia} DR3 ID 3456522854428709632. The inferred RV semi-amplitude is 21 km s$^{-1}$. In both panels, we plot fitted RV curves corresponding to 100 random samples of ($\gamma$, $K$) from the posterior. Similar fits for all 14 variables are shown in Figure~\ref{fig:all_rvs}.}
\label{fig:typical_rv}
\end{figure*}

We show two examples of radial velocity curves, along with their corresponding observations (color-coded by instrument used), in Figure \ref{fig:typical_rv}. The left panel shows an RV curve with a typical number of observations, while the right panel shows one of the targets for which we obtained the largest number of observations. Figure \ref{fig:all_rvs} in the appendix shows similar RV curves for all the objects in our follow-up sample. In general, we infer relatively small RV semi-amplitudes; the median value is about $17$ km s$^{-1}$, and the maximum value is about $61$ km s$^{-1}$.  A few outlier data points are most likely due to systematic errors due to luminous secondaries changing the shape of spectral lines between epochs.

\subsection{Companion Masses}
\label{ssec:companions}

We estimated the primary star mass, $M_1$, from its $G$-band absolute magnitude using the empirical $G$-band mass-luminosity relationship for main-sequence stars from \citet{janssens_2022}. Then, given the orbital period $P$, semi-amplitude $K$, and an assumed inclination $i$, the companion mass $M_2$ can be found by numerically solving the following equation:

\begin{equation}
    K = \left(\frac{2 \pi G}{(M_1 + M_2)^2 P}\right)^{1/3} M_2 \sin{i}
\end{equation}

\begin{figure}
    \includegraphics[width=\columnwidth]{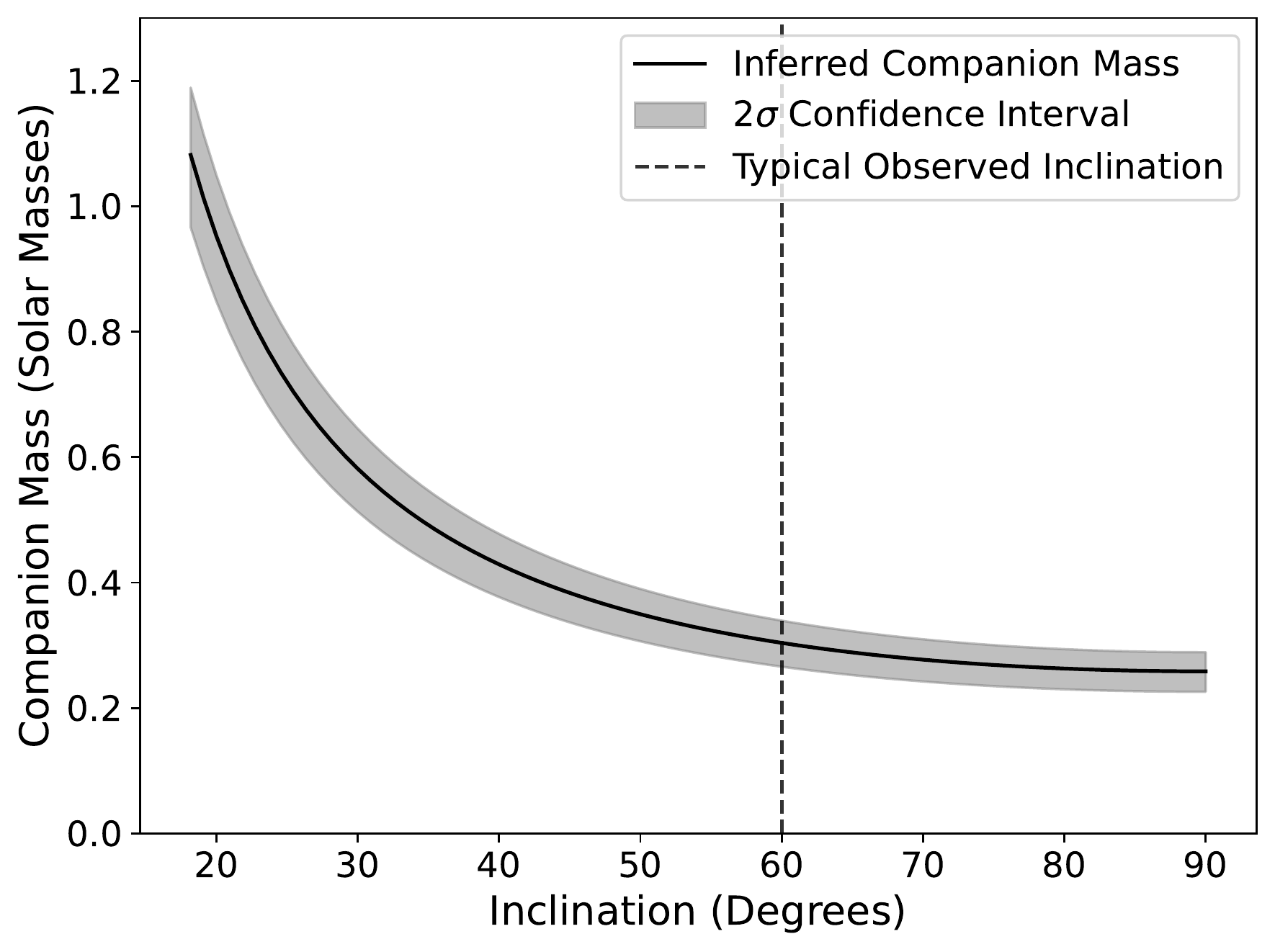}
    \caption{Dependence of derived companion mass (given the observed period and RV semi-amplitude) on inclination for ellipsoidal variable candidate with \textit{Gaia} DR3 ID 3457961423656077056. Although the companion mass could in principle be large for a sufficiently face-on inclination, a typical inclination of 60$^{\circ}$ implies a companion mass of only $0.3 ~ M_{\odot}$.}
    \label{fig:inclination_mass}
\end{figure}

We infer minimum companion masses by assuming an inclination of 90 degrees. For randomly oriented orbits, the expected median inclination is 60 degrees. The typical inclination for objects in our sample may differ from this value: relatively edge-on orbits may be favored by the selection of large variability amplitudes, but the most edge-on orbits may be excluded because (for luminous companions) they would show eclipses. Figure \ref{fig:inclination_mass} shows how the inferred companion mass for a typical object in our sample depends on the assumed inclination for a fixed value of $K$. The error bars represent a 2$\sigma$ confidence interval derived based on the uncertainty in the primary mass. We observe that the inferred companion mass increases with decreasing inclination, with the inferred $M_2$ at a typical inclination of $60^{\circ}$ only differing slightly from the minimum inferred mass. Only for nearly face-on inclinations, which are geometrically improbable, does the actual companion mass exceed the minimum by more than a factor of two. In fact, while the observed RV semi-amplitude would imply a companion mass higher than $1 ~ M_{\odot}$ for inclination $i < 25^{\circ}$, such a low inclination would also result in a smaller photometric variability amplitude than exhibited by the observed systems (see simulations in Section \ref{subsec:phoebe_models}).

\begin{figure*}
    \includegraphics[width=\textwidth]{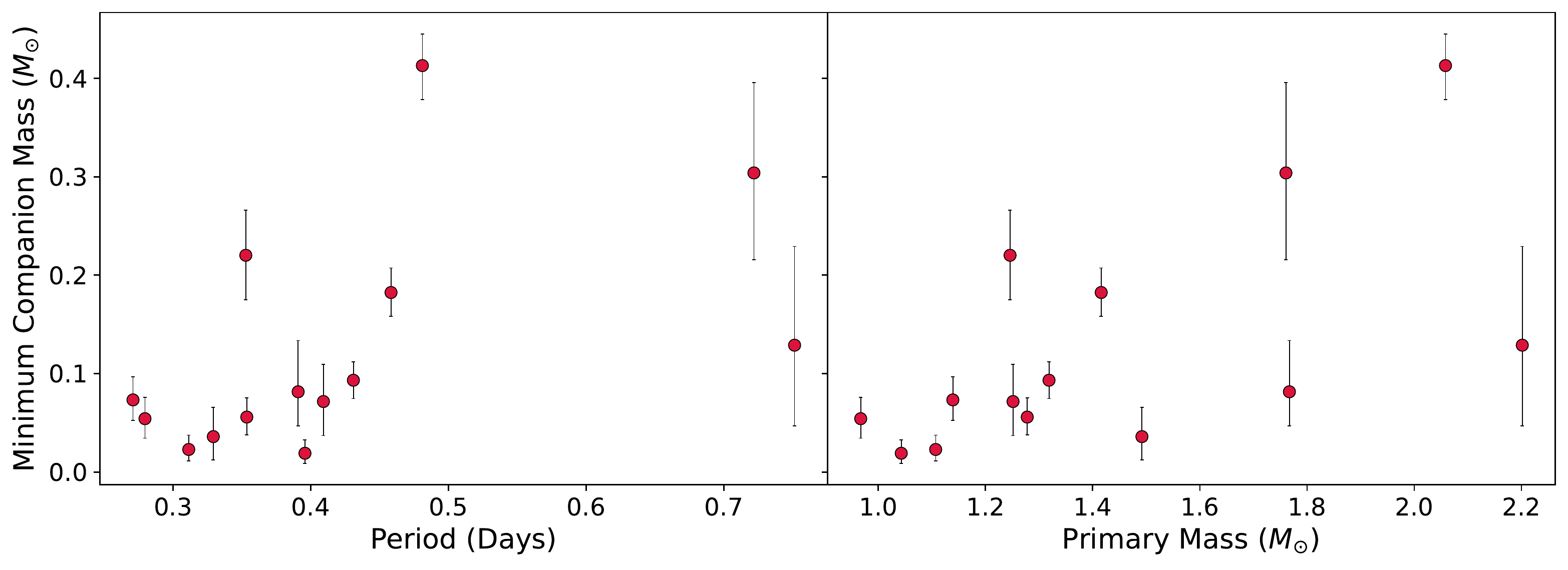}
    \caption{Left Panel: Derived minimum companion mass as a function of period. Right Panel: Derived companion mass as a function of primary mass. Our assumed inclination is $90^{\circ}$. Most of our inferred companion masses are near $0.1 ~ M_{\odot}$ -- much lower than expected for any compact object companion. We do not observe a strong trend with either orbital period or primary mass.}
    \label{fig:summary_fig}
\end{figure*}

To provide a summary of our results in Table \ref{tab:results}, we plot the inferred companion mass versus period and primary mass for an inclination of $90^{\circ}$ in Figure \ref{fig:summary_fig}. There is a weak trend of increasing companion mass with longer periods and higher primary masses. Our median inferred companion mass is about $0.08$ solar masses; if we assume an inclination of $60^{\circ}$, we infer a median companion mass of $0.09$ solar masses instead. The implications of this are discussed in Section \ref{sec:discussion}.

\section{Discussion} \label{sec:discussion}

Our spectroscopic follow-up of the most promising ellipsoidal variables in the sample selected from {\it Gaia} DR3 by \citet{gomel_gaia_2022} revealed a population of binaries with mass ratios much less than unity, contrary to expectations. Hence, we must reject the hypothesis that these binaries are ellipsoidal variables hosting massive dark companions. We now consider four alternative explanations for the objects' photometric variability.

\begin{enumerate}
    \item The sources are are indeed ellipsoidal variables with dark companions, but the companions have low masses ($M_2\sim 0.2\,M_{\odot}$).  We consider this unlikely: it is not clear what the low-mass dark secondaries would actually be, and their mass ratios are in any case not large enough to explain the observed variability through tidal deformation of the primary alone. 
    \item The sources are actually detached binaries with a low-mass, luminous secondaries (i.e., main-sequence stars). 
    \item The sources are are actually contact binaries (meaning that two luminous components share a common envelope), with an extreme mass ratio, and possibly with spots on either component. 
    \item The sources are not binaries at all, but rotational or pulsating variables whose light curves just happen to mimic ellipsoidal variability. 
\end{enumerate}

We explore the expected light curves and RV amplitudes for plausible luminous binary scenarios below using PHOEBE \citep{Wilson_Devinney_1971, Prvsa_Zwitter_2005, conroy_2020}, which models the light curves and RVs of close binaries. 

\subsection{PHOEBE Light Curve Models}
\label{subsec:phoebe_models}

\begin{figure*}
\includegraphics[width=0.9\textwidth]{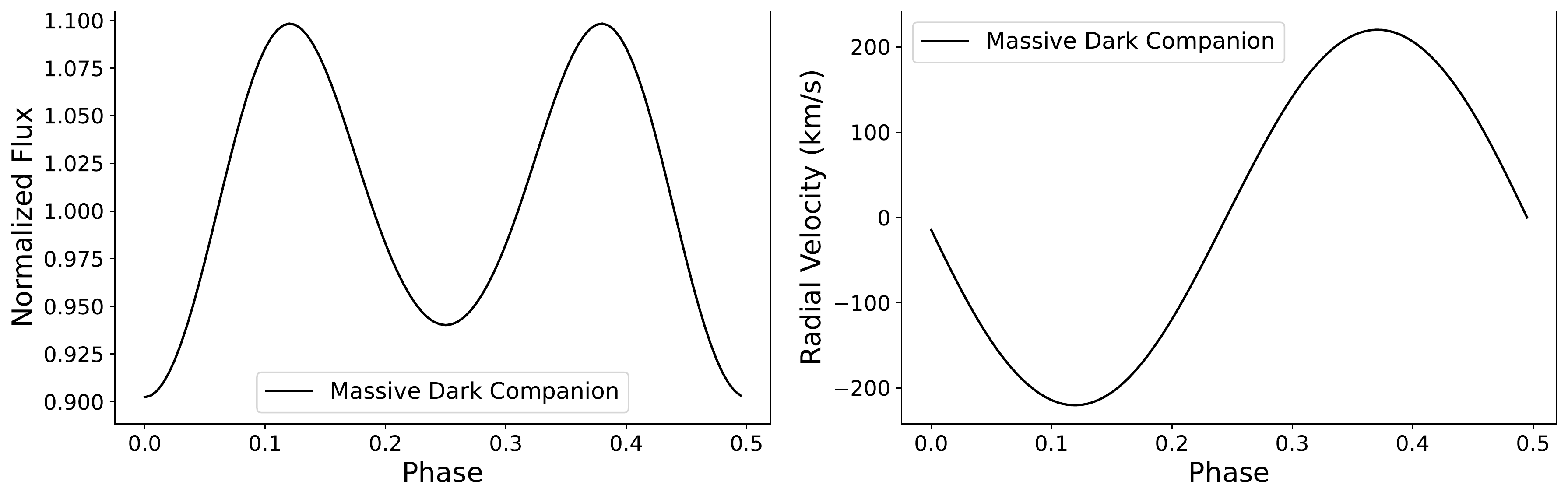}
\includegraphics[width=0.9\textwidth]{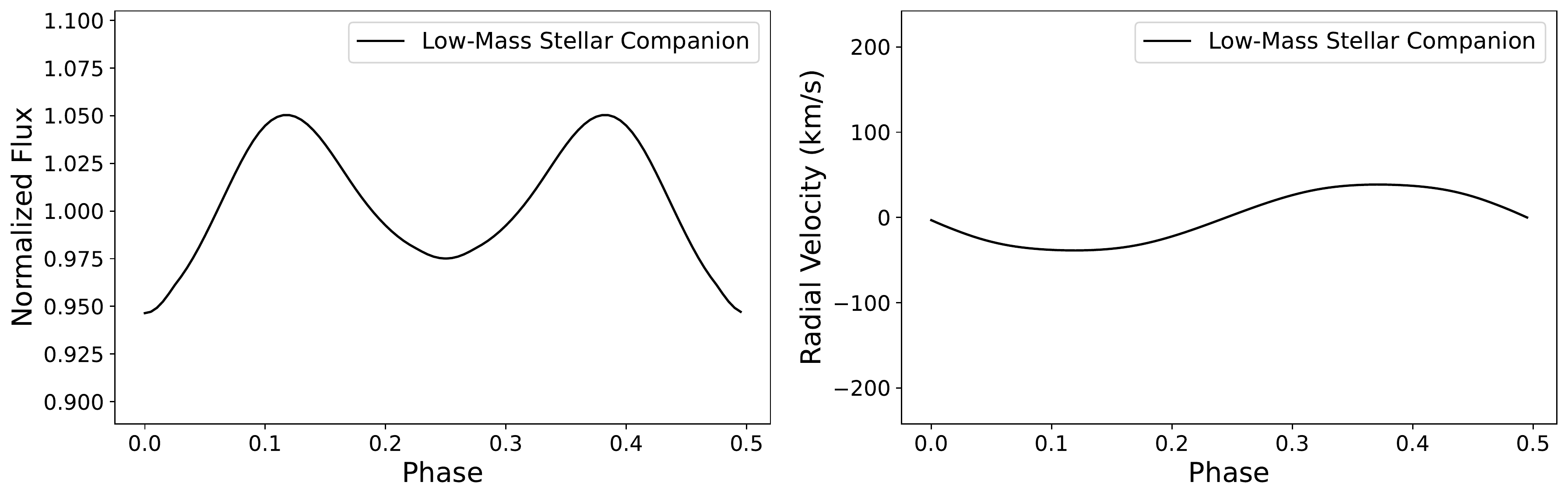}
\includegraphics[width=0.9\textwidth]{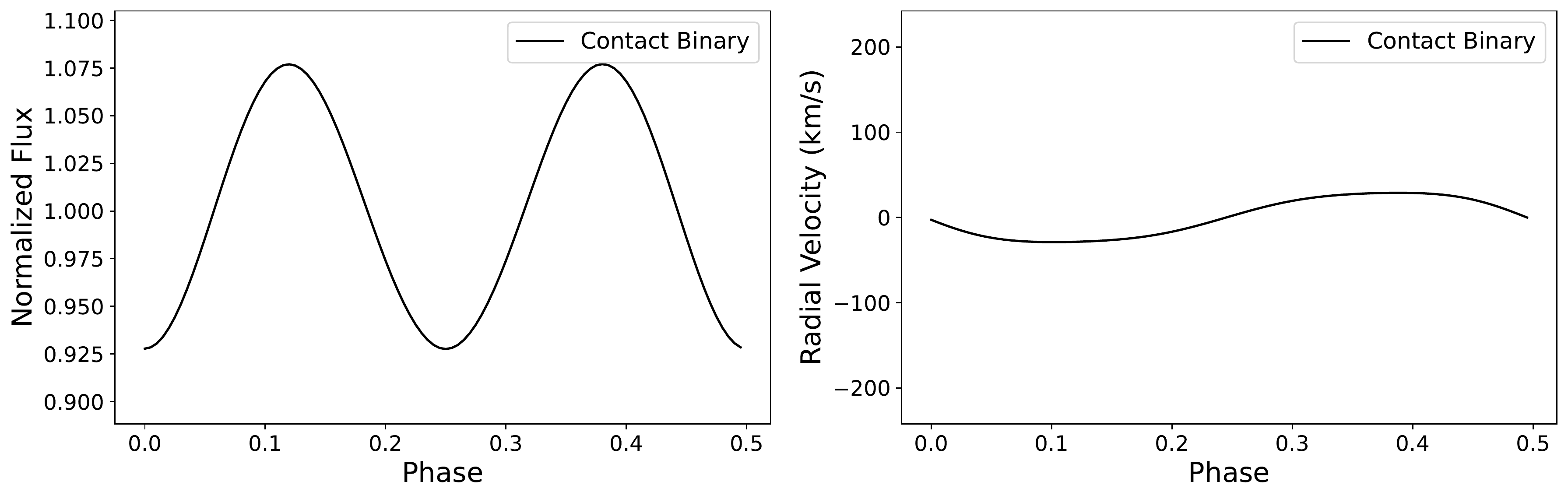}
\includegraphics[width=0.9\textwidth]{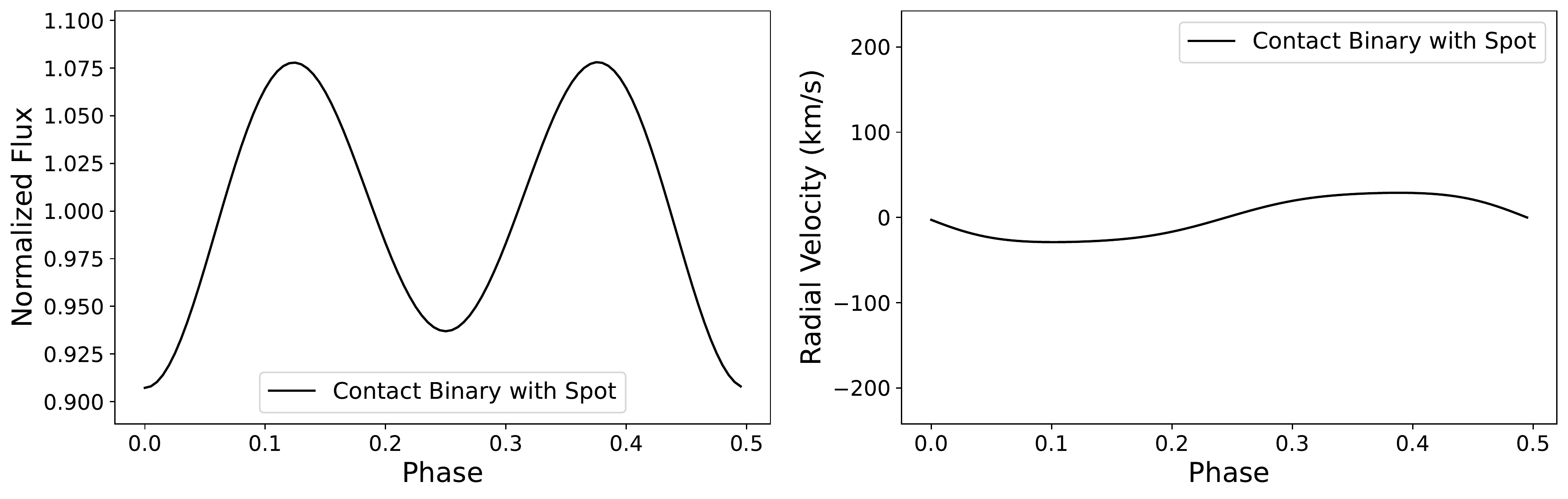}
\caption{Row One: Simulated light curve and RV curve for a $q = 2.0$ ellipsoidal variable with a massive dark companion (assuming an inclination of 60$^{\circ}$). Row Two: Simulated light curve and RV curve for a detached binary with $q = 0.2$ (assuming an inclination of 60$^{\circ}$). Row Three: Simulated light curve and RV curve for a contact binary with $q = 0.2$ (assuming an inclination of 45$^{\circ}$). Row Four: Simulated light curve and RV curve for a spotted contact binary with $q = 0.2$ (assuming an inclination of 45$^{\circ}$). We find that only a contact binary with a spot can reproduce the large variability amplitude, ``ellipsoidal'' light curve shape, and low RV semi-amplitude characteristic of our observations.}
\label{fig:phoebe_comparison}
\end{figure*}

\begin{figure*}
\centering
\begin{subfigure}{.49\textwidth}
  \centering
  \includegraphics[width=\columnwidth]{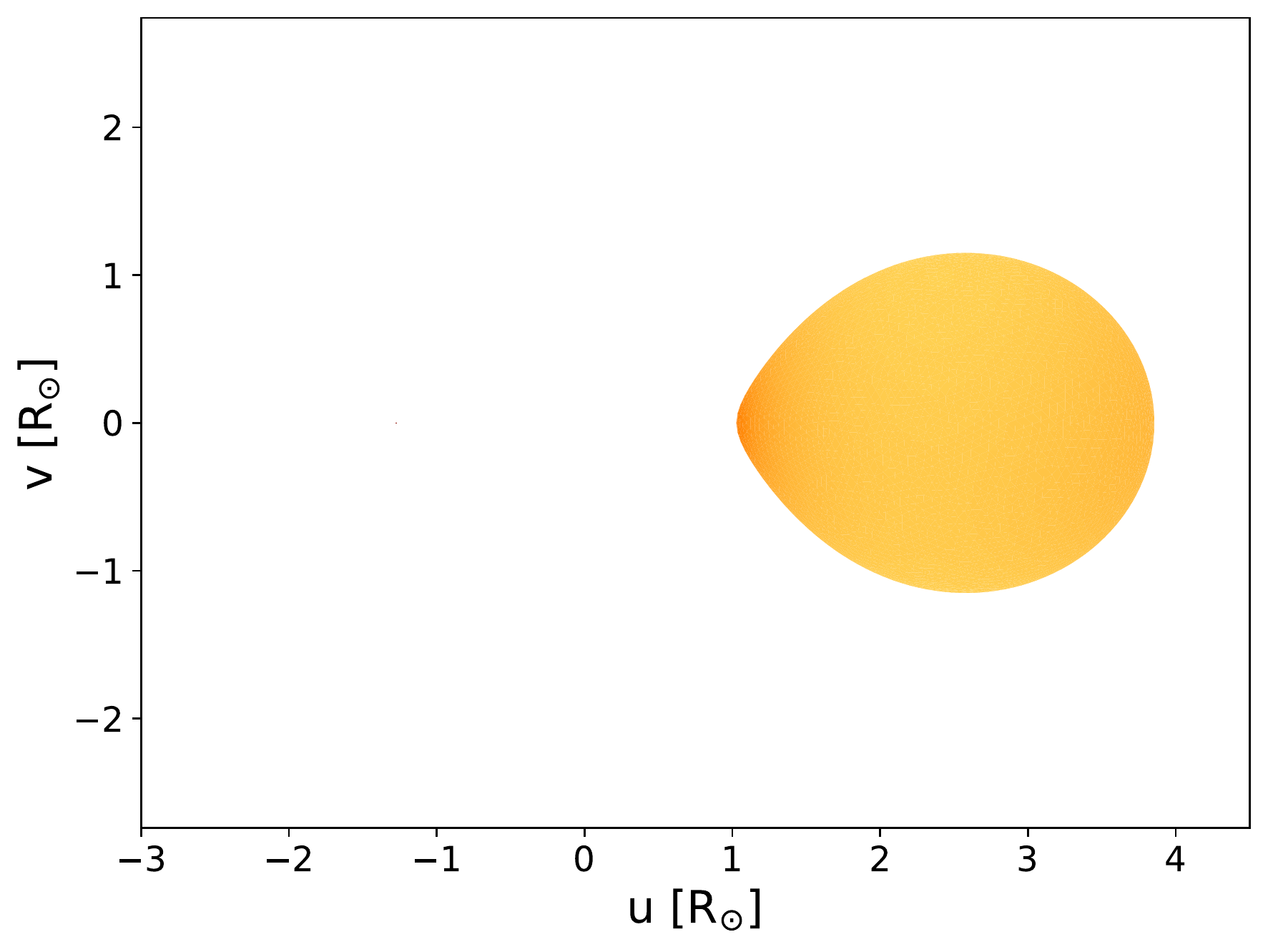}
\end{subfigure}
\begin{subfigure}{.49\textwidth}
  \centering
  \includegraphics[width=\columnwidth]{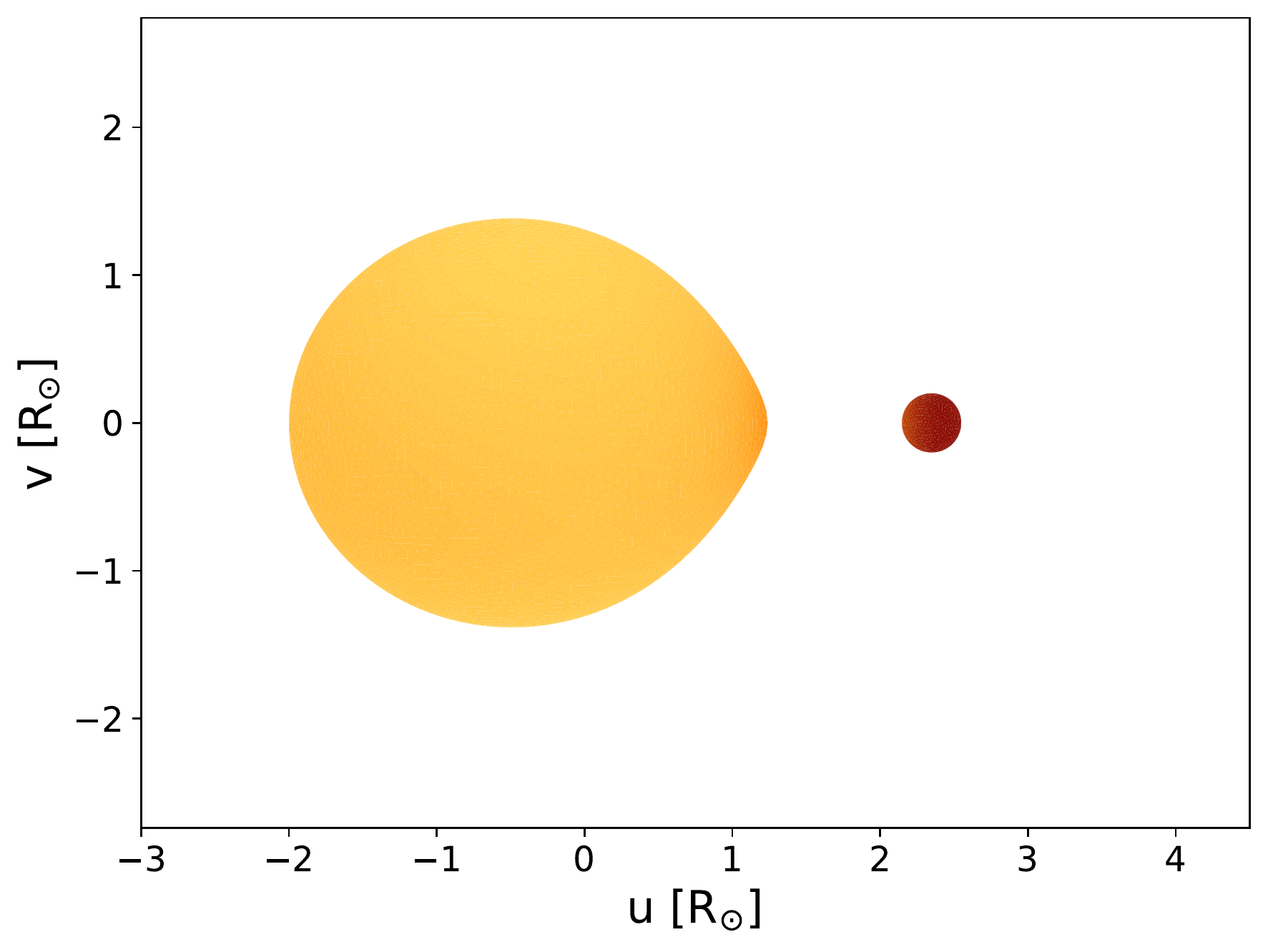}
\end{subfigure}
\vskip\baselineskip
\begin{subfigure}{.49\textwidth}
  \centering
  \includegraphics[width=\columnwidth]{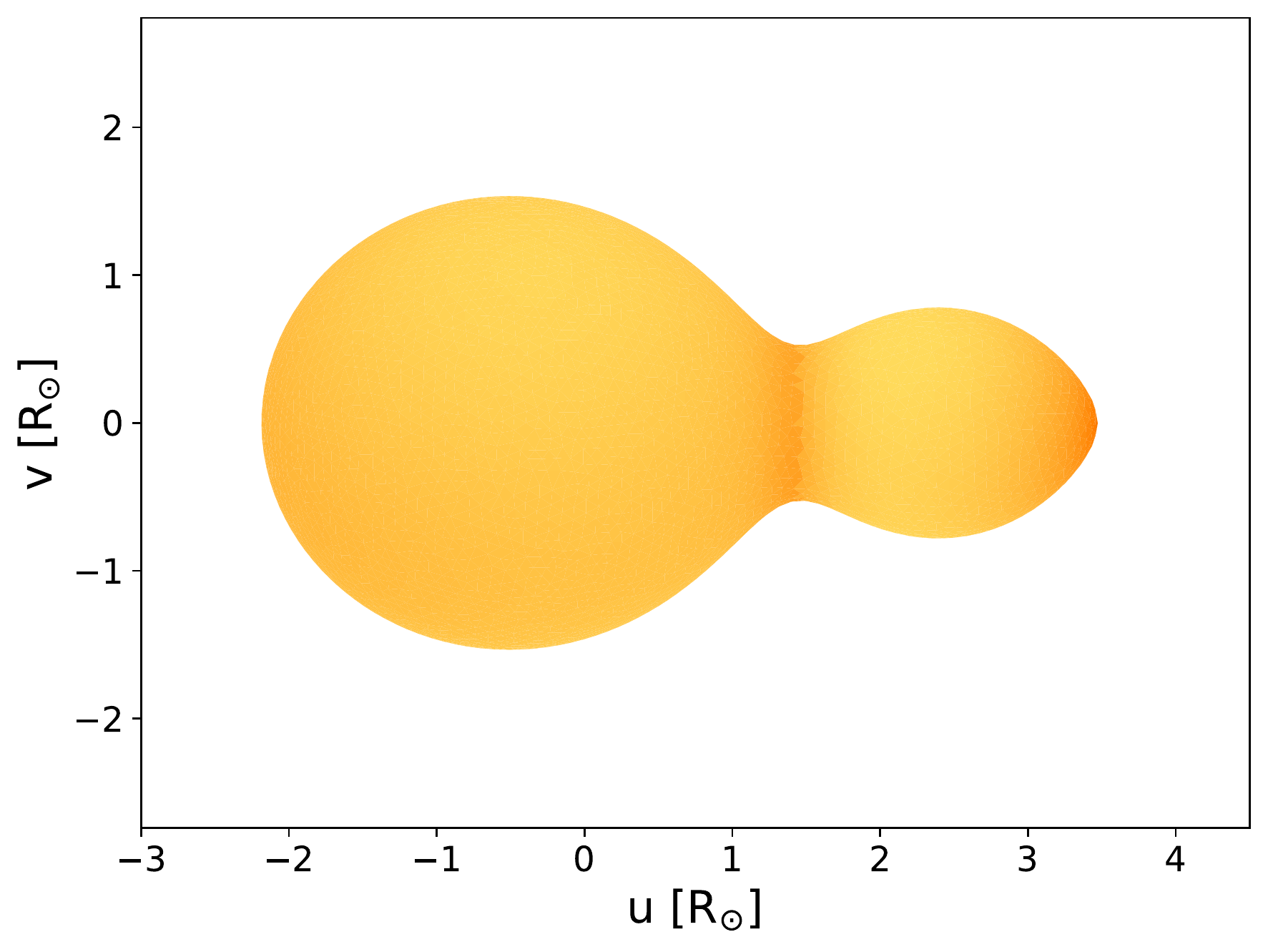}
\end{subfigure}
\begin{subfigure}{.49\textwidth}
  \centering
  \includegraphics[width=\columnwidth]{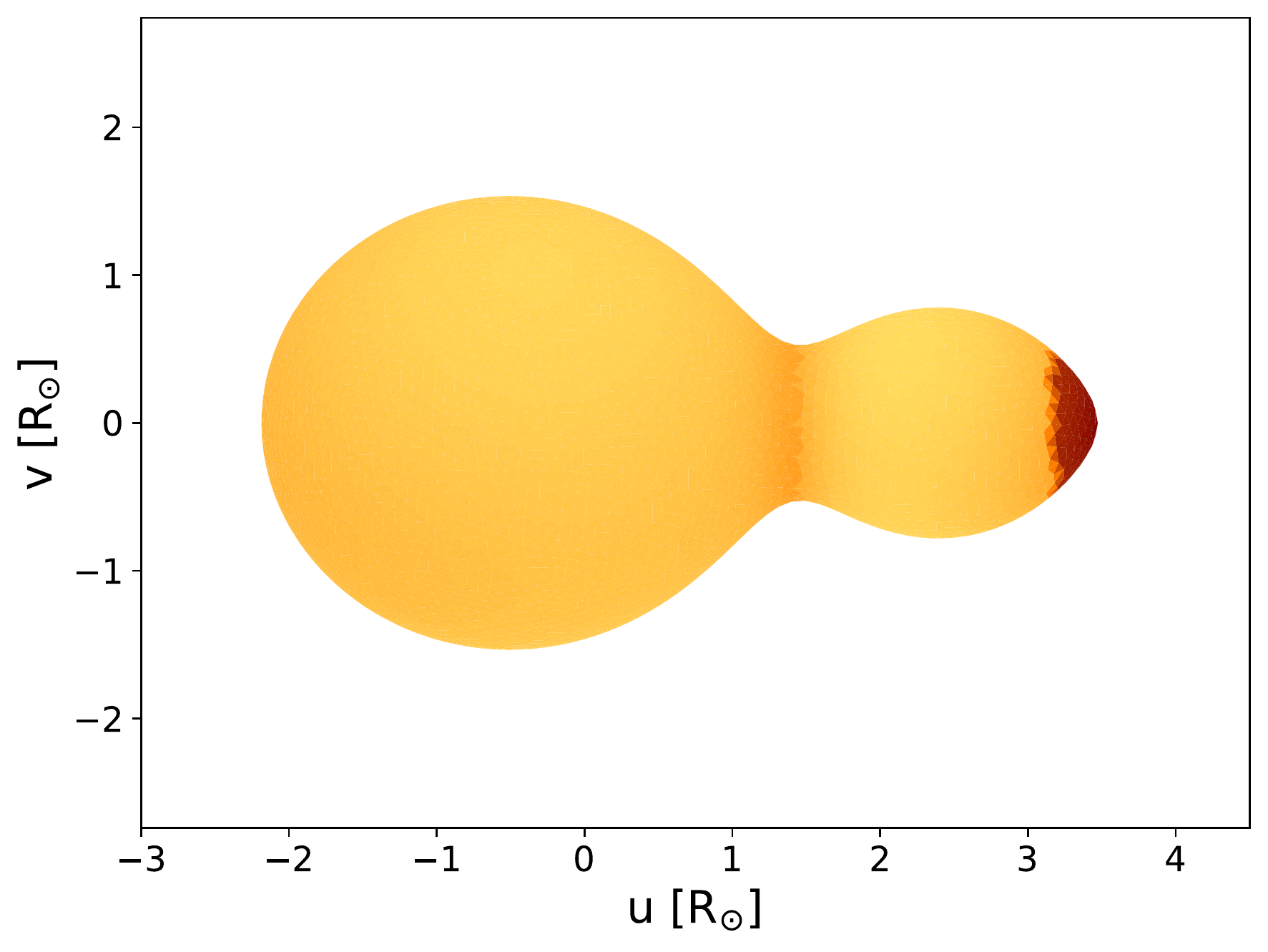}
\end{subfigure}
\caption{Top Row: Visualization of a $q = 2.0$ ellipsoidal variable with a massive dark companion and a $q = 0.2$ detached binary with a low mass secondary (assuming an inclination of 60$^{\circ}$).  Bottom Row: Visualization of a $q = 0.2$ extreme mass ratio contact binary with and without a spot (assuming an inclination of 45$^{\circ}$). In both cases, the mesh is colored by effective temperature.}
\label{fig:phoebe_mesh}
\end{figure*}

We used PHOEBE to calculate model light curves and RV curves  by specifying the mass ratio, orbital period, inclination, semi-major axis, primary and secondary radii, and primary and secondary temperatures as input parameters. The semi-major axis was self-consistently determined based on the primary mass, secondary mass, and orbital period. We investigated four different models, all with a primary mass of $1 ~ M_{\odot}$, a primary temperature of $6000$ K, and an orbital period of 0.5 days. For each possible scenario, we adjusted parameters such as the inclination, component radii, and spot morphology (where applicable) to produce light curves similar to the observed light curves in our sample. We show predicted light curves and primary RVs for four models in Figure~\ref{fig:phoebe_comparison}; the corresponding meshes are shown in Figure~\ref{fig:phoebe_mesh}. The models are as follows:

\begin{enumerate}
    \item A default ellipsoidal binary with a secondary mass of $2 ~ M_{\odot}$ (i.e.\ $q = 2.0$) at an observed inclination of $60^{\circ}$. In this case, we set the radius of the primary to 0.99 times the effective Roche lobe radius. We also set the radius of the secondary to a vanishingly small value to mimic an unseen massive companion. This is the scenario corresponding to a true BH/NS companion. It predicts a light curve with uneven minima and photometric variability on half of the orbital period. The corresponding radial velocity curve has a semi-amplitude greater than $200$ km s$^{-1}$, much larger than any of our observed binaries.
    \item A detached binary with a low-mass luminous secondary of mass $0.2 ~ M_{\odot}$ (i.e.\ $q = 0.2$) at an observed inclination of $60^{\circ}$. In this case, we set the radius of the primary to 0.99 times the effective Roche lobe radius. We also set the radius of the secondary to $0.2 ~ R_{\odot}$ and the temperature of the secondary to $4000$ K, as appropriate for a low-mass main sequence star. While the light curve still displays the typical ellipsoidal features described above, it has a smaller variability amplitude. On the other hand, the corresponding radial velocity curve has a semi-amplitude less than $50$ km s$^{-1}$, in accordance with our observations.
    \item A contact binary with a low-mass luminous secondary of mass $0.2 ~ M_{\odot}$ (i.e.\ $q = 0.2$) at an observed inclination of $45^{\circ}$. In this case, we set the radius of the primary to 1.09 times the effective Roche lobe radius. While we did not specify the radius of the secondary (as it is no longer a free parameter), we did set the temperature of the secondary to $6000$ K (as we expect the components to be in thermal equilibrium). In this case, the radial velocity curve maintains a small semi-amplitude, but the light curve displays a lack of unequal minima.
    \item A spotted contact binary with a low-mass luminous secondary of mass $0.2 ~ M_{\odot}$ (i.e.\ $q = 0.2$) at an observed inclination of $45^{\circ}$. Once again, we set the radius of the primary to 1.09 times the effective Roche lobe radius, and the temperature of the secondary to $6000$ K. However, we also added a spot of temperature $4500$ K and an angular radius of 30$^{\circ}$ at a co-latitude of $90^{\circ}$ and a longitude of $180^{\circ}$ with respect to the spin axis. The location of the spot was chosen to produce uneven minima in the observed light curve. Of course, this is not the only configuration of spots that could produce apparent ellipsoidal variability: a spot of a different angular radius and effective temperature or the addition of another spot with different physical characteristics on the opposing end of the binary would also produce observed ellipsoidal modulation, albeit with a different variability amplitude. However, we decided not to consider scenarios involving multiple spots for the sake of simplicity.
\end{enumerate}

We observe that our base case hypothesis (i.e.\ an ellipsoidal variable with $q = 2.0$) produces a typical ellipsoidal light curve with a large RV semi-amplitude. The light curve matches our observations, but the radial velocity curve does not. On the other hand, both a detached binary with $q = 0.2$ and a contact binary with $q = 0.2$ produce observed RV curves with small semi-amplitudes that match our observations. In the scenario of a contact binary with a spotted secondary, the light curve still displays a characteristic difference in minima that matches our observations; this is because the spot mimics the effects of gravity darkening in an ellipsoidal binary (a ``reverse'' O'Connell effect, see \citealt{oconnell_1951}). While the spotted contact binary seems to match our observed data the best, it is also the most complicated model. 

\subsubsection{Extreme mass ratio contact binaries}

Due to mass transfer, magnetic braking, and thermal relaxation oscillations, the mass ratio of a contact binary evolves towards small values until it reaches a critical threshold \citep{eggleton_2012}. At this threshold, the Darwin instability will cause a runaway loss of angular momentum that leads to a coalescence \citep{eggleton_2012}. Recent studies of observed samples of contact binaries have modeled the mass ratio distribution of contact binaries as a function of orbital period, and have found that extreme mass ratio contact binaries (i.e. $q \sim 0.2$) are not uncommon at the typical periods ($P \sim 0.3$ days) in our sample \citep[][]{kobulnicky_2022, pesta_2022}. We thus suspect that (spotted) contact binaries are among the most common false-positives in the {\it Gaia} sample.

\subsubsection{Evidence of spots in the phased light curves}

\begin{figure*}
\centering
\begin{subfigure}{.49\textwidth}
  \centering
  \includegraphics[width=\columnwidth]{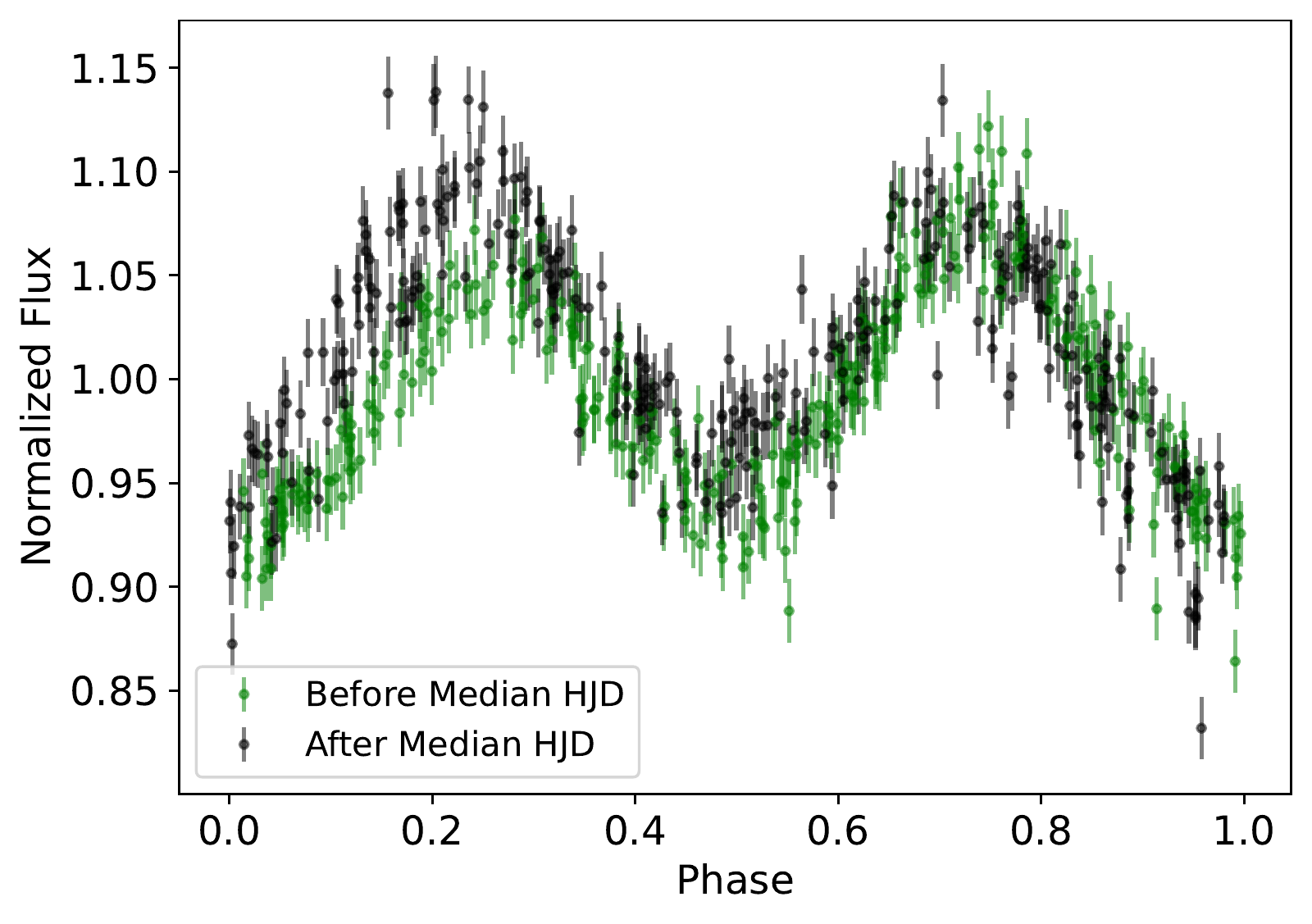}
\end{subfigure}
\begin{subfigure}{.49\textwidth}
  \centering
  \includegraphics[width=\columnwidth]{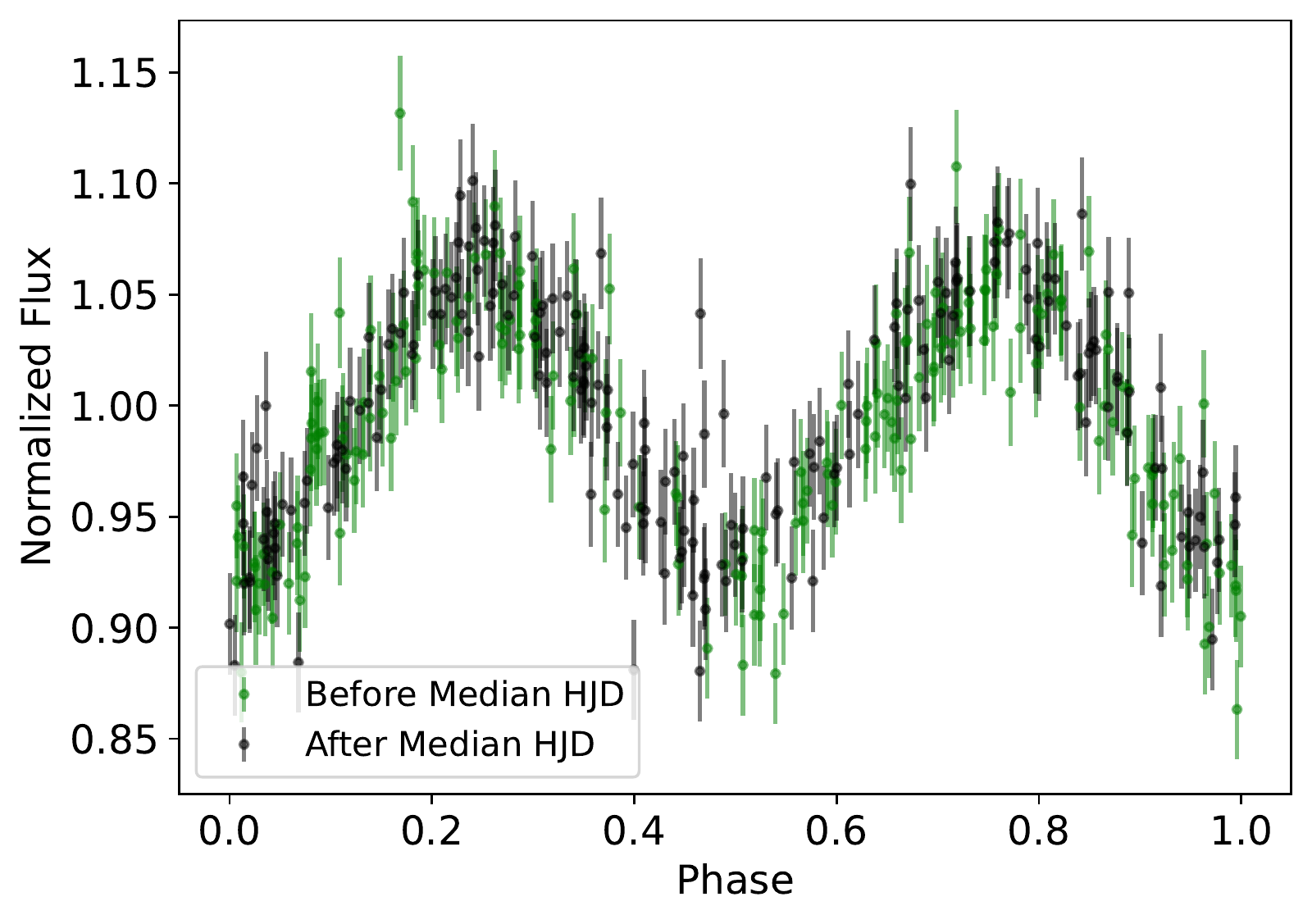}
\end{subfigure}
\caption{Phased ZTF $g$-band light curves for two objects from our sample. In both panels, we split the light curve into two sections, each corresponding to $\approx 2$ years. In the left panel (\textit{Gaia} DR3 ID 1342190925113500416), there is clear evolution of the light curve morphology between the two sections, a telltale sign of spot evolution. In the right panel (\textit{Gaia} DR3 ID 4318115181190923392), the light curve is stable. This does not rule out spots, but it suggests that there was little spot evolution over over $\sim$4 years. }
\label{fig:long_timescale}
\end{figure*}

To check for long-timescale variations in the ZTF light curves that could be caused by spots moving on stellar surfaces, we divided the data for each target in our sample into two segments of equal length and visually searched for changes in phase, amplitude, or shape between the two segments. In a few sources, we did find evidence of the phased light curve shape evolving in a manner consistent with spot activity. A majority of the objects in the sample did not display convincing evidence of light curve shape evolution. We provide examples of ZTF $g$-band light curves with and without long-timescale variations in Figure \ref{fig:long_timescale}.

The fact that some objects in our sample show evidence of spot evolution supports the hypothesis that spots contribute to the observed variability in at least some of the sample. Spot evolution timescales in close binaries can span decades \citep[e.g.][]{Eaton1979}, so the lack of detected light curve evolution in most of the targets does not preclude spots. 

\subsection{Single Star Scenarios}

We now consider possible scenarios in which the sources are not binaries at all, but just single variable stars with light curves that serendipitously resemble ellipsoidal variables. One such plausible case is a rotating single-star with spots that generate photometric variability on the rotational period. Another potential scenario is a pulsating variable star; however, pulsations are unlikely due to the observed rapid rotation ($v \sin{i} \sim 100 $ km s$^{-1}$), and lack of known astrophysical pulsations at these timescales and amplitudes in this part of the CMD. More convincingly, the fact that most of the objects in the sample are above the main-sequence suggests binarity, as it is improbable that a single star could produce the observed over-luminosity. In the end, it is important to note that there could be more than one type of contaminant in the sample of candidate ellipsoidal variables.

\subsection{Completeness of our follow-up}

We observed 14 candidates on the main sequence of the extinction-corrected \textit{Gaia} CMD, prioritizing bright sources with large photometrically-inferred minimum mass ratios. While this represents a small fraction of the 6306 candidates selected by \citet{gomel_gaia_2022}, we expect that the false-positive rate in our spectroscopic follow-up sample is {\it lower} than in the full \citet{gomel_gaia_2022} sample because we excluded (a) sources above the main-sequence, which are likely luminous binaries and (b) sources with ZTF light curves displaying systematic deviations from ellipsoidal variability. 
Only one of the objects in our spectroscopic sample (\textit{Gaia} DR3 ID 4310698008860121216) is among the 262 candidates identified by \citet{gomel_gaia_2022} that had inferred mass ratios larger than unity at the $1\sigma$ level. However, the majority of these 262 candidates lie above the main sequence in the extinction-corrected \textit{Gaia} CMD, and hence were excluded from our final sample. In any case, the minimum mass ratios inferred from the RV curves for the 14 candidates in our sample do not match the $\hat{q}_{\min}$ values inferred from the variability amplitudes of the light curves. There is little reason to expect the photometrically-inferred minimum mass ratios to be any more reliable for the sources with the highest $\hat{q}_{\min}$ values, particularly since sources with genuinely dark companions are astrophysically rare. Based on the ``success rate'' of 0/14 within our spectroscopic follow-up sample, a conservative lower limit on the false-positive rate for the full catalog is $\gtrsim 90\%$. 

\subsection{Challenges of using the mMMR in searches for compact object companions}

The method suggested by \citet{gomel_faigler_mazeh_2021} is intended to search for objects that are close to completely filling their Roche lobe (i.e.\ $f \gtrsim 0.90$). However, this is a short-lived phase in the evolution of most binaries compared to evolutionary phases where both components are well within their Roche lobes or one component is completely Roche-lobe filling \citep[e.g.][]{Warner2003}. Hence, the population of observed binaries should primarily consist of binaries that are well detached (and not identifiable as hosting massive dark companions via the mMMR) or systems where the primary is already overflowing its Roche lobe. In fact, if a stellar mass BH companion was present in the latter case, we would expect such systems to be transient or persistent X-ray sources (though sources with long outburst recurrence times might not be easily detectable).

As a result, the central weakness of the method proposed by \citet{gomel_faigler_mazeh_2021} is that selection effects will cause a sample in which light curves display ellipsoidal modulation to be dominated by false positives. The most likely contaminants are close luminous binaries, which are more common than ellipsoidal variables with dark companions. In fact, among solar-type stars, about one out of every 200 sources is a binary with a period less than 1 day \citep[e.g.][]{el_badry_2022c}. While the light curves of such binaries will often be distinguishable from those of ellipsoidal variables, their greater numbers and the ability of spots to distort light curves in nearly arbitrary ways make them a significant contaminant. However, single stars with rotational or pulsational variation can also serve as potential contaminants. While the sample curated by \citet{gomel_gaia_2022} could still contain stellar-mass BHs and NSs, a wider characterization of the sample (i.e.\ with a high degree of completeness) is required to further determine the sensitivity and specificity of their method.

\section{Conclusion} \label{sec:conclusion}

We performed spectroscopic follow-up of 14 of the most promising variables in the \textit{Gaia} DR3 sample of ellipsoidal variables curated by \citet{gomel_gaia_2022}. In previous work, \citet{gomel_faigler_mazeh_2021} demonstrated that, under the assumption that all variability is indeed due to tidal deformation, a stringent lower limit on the mass ratio of an ellipsoidal variable can be inferred that is robust to assumptions about the mass, radius, and evolutionary state of the component stars. They termed this lower limit the ``modified minimum mass ratio'' (mMMR or $\hat{q}_{\min}$). Of the candidates in our sample, 12 had $\hat{q}_{\min} > 0.8$ and 10 had $\hat{q}_{\min} > 1.0$; however, we do not find any promising compact object companions. Our main conclusions are as follows:

\begin{itemize}
    \item We selected our 14 targets based on their optical light curves, whose amplitude variability (when attributed entirely to a massive unseen companion) suggested that the secondary was a compact object (see Figure \ref{fig:all_lcs}). All of these targets had apparent \textit{Gaia} $G$-band magnitude $\leq 17$, well-constrained distances, available 3D dust map extinctions, and CMD positions on the main sequence (see Figure \ref{fig:sample}). We obtained spectra for each of these targets at at least 2 epochs, using the DBSP spectrograph as a primary tool (and the LRIS and ESI spectrographs at Keck as available). From the fitted semi-amplitudes of the RV curves (see Figure \ref{fig:all_rvs}), we derive minimum companion masses of $\leq 0.5 ~ M_{\odot}$ in all cases, rendering it unlikely that any of these binary systems host neutron stars or black holes (see Figure \ref{fig:summary_fig}).
    \item We use PHOEBE modeling to explore detached systems with low-mass luminous secondaries and (spotted) contact binaries as alternative hypotheses for how binary systems with low mass ratios could produce the observed light curves (see Figures \ref{fig:phoebe_comparison} and \ref{fig:phoebe_mesh}). We find that an extreme mass ratio contact binary with a spotted secondary can reproduce both an RV curve with a small semi-amplitude and a light curve with apparent ellipsoidal modulation.
    \item Since we selected candidates based on the position of the primary component relative to the main sequence on the \textit{Gaia} CMD (which rules out Algol-type binaries) (see Figure \ref{fig:sample}), we conclude that the most common failure mode was probably mistaking a (spotted) contact binary for an ellipsoidal variable. We note that resource constraints restricted our investigation to a small subset of the original sample; there may still be stellar mass BHs in the dataset provided by \citet{gomel_gaia_2022}. However, a significant fraction of their candidates can be eliminated right away based on their CMD position above the main sequence.
    \item Thus, further characterization of this sample (i.e.\ with a higher degree of completeness) is necessary to determine whether any BH/NS needles are hiding in the luminous binary haystack. All-sky, multi-epoch spectroscopic surveys such as SDSS-V \citep{kollmeier_2017}, which are capable of optical multi-object fiber spectroscopy, will be ideally suited to carry out this task. Spectroscopic follow-up of ellipsoidal binaries remains a promising avenue to understand the progenitors of low-mass X-ray binaries and characterize the population of dormant stellar mass black holes, but it must be approached with the assumption that a majority of light-curve selected candidates are false positives. 
\end{itemize}

\section*{Acknowledgements}

We thank the referee for a constructive report. We are grateful to Roy Gomel and Tsevi Mazeh for providing an early copy of their candidate list, Tsevi Mazeh for comments on an early draft of this manuscript, and the ZTF Variable Star Group for useful comments and discussions. This work made use of Astropy, a community-developed core Python package and an ecosystem of tools and resources for astronomy \citep{2018AJ....156..123A}.

This work is based on observations obtained with the Samuel Oschin Telescope 48-inch and the 60-inch Telescope at the Palomar Observatory as part of the Zwicky Transient Facility project. ZTF is supported by the National Science Foundation under Grant
No. AST-2034437 and a collaboration including Caltech, IPAC, the Weizmann Institute for Science, the Oskar Klein Center at Stockholm University, the University of Maryland, Deutsches Elektronen-Synchrotron and Humboldt University, the TANGO
Consortium of Taiwan, the University of Wisconsin at Milwaukee, Trinity College Dublin, Lawrence Livermore National Laboratories, and IN2P3, France. Operations are conducted by COO, IPAC, and UW.

This work has made use of data from the European Space Agency (ESA) mission {\it Gaia} (\url{https://www.cosmos.esa.int/gaia}), processed by the {\it Gaia} Data Processing and Analysis Consortium (DPAC, \url{https://www.cosmos.esa.int/web/gaia/dpac/consortium}). Funding for the DPAC has been provided by national institutions, in particular
the institutions participating in the {\it Gaia} Multilateral Agreement.

%%%%%%%%%%%%%%%%%%%%%%%%%%%%%%%%%%%%%%%%%%%%%%%%%%
\section*{Data Availability}

Data used in this study are available upon request from the corresponding author.



\bibliographystyle{mnras}
\bibliography{bibliography} % if your bibtex file is called example.bib

\begin{thebibliography}{}
\makeatletter
\relax
\def\mn@urlcharsother{\let\do\@makeother \do\$\do\&\do\#\do\^\do\_\do\%\do\~}
\def\mn@doi{\begingroup\mn@urlcharsother \@ifnextchar [ {\mn@doi@}
  {\mn@doi@[]}}
\def\mn@doi@[#1]#2{\def\@tempa{#1}\ifx\@tempa\@empty \href
  {http://dx.doi.org/#2} {doi:#2}\else \href {http://dx.doi.org/#2} {#1}\fi
  \endgroup}
\def\mn@eprint#1#2{\mn@eprint@#1:#2::\@nil}
\def\mn@eprint@arXiv#1{\href {http://arxiv.org/abs/#1} {{\tt arXiv:#1}}}
\def\mn@eprint@dblp#1{\href {http://dblp.uni-trier.de/rec/bibtex/#1.xml}
  {dblp:#1}}
\def\mn@eprint@#1:#2:#3:#4\@nil{\def\@tempa {#1}\def\@tempb {#2}\def\@tempc
  {#3}\ifx \@tempc \@empty \let \@tempc \@tempb \let \@tempb \@tempa \fi \ifx
  \@tempb \@empty \def\@tempb {arXiv}\fi \@ifundefined
  {mn@eprint@\@tempb}{\@tempb:\@tempc}{\expandafter \expandafter \csname
  mn@eprint@\@tempb\endcsname \expandafter{\@tempc}}}

\bibitem[\protect\citeauthoryear{{Astropy Collaboration} et~al.,}{{Astropy
  Collaboration} et~al.}{2018}]{2018AJ....156..123A}
{Astropy Collaboration} et~al., 2018, \mn@doi [\aj] {10.3847/1538-3881/aabc4f},
  \href {https://ui.adsabs.harvard.edu/abs/2018AJ....156..123A} {156, 123}

\bibitem[\protect\citeauthoryear{{Bellm} et~al.,}{{Bellm}
  et~al.}{2019}]{bellm_2019}
{Bellm} E.~C.,  et~al., 2019, \mn@doi [\pasp] {10.1088/1538-3873/aaecbe}, \href
  {https://ui.adsabs.harvard.edu/abs/2019PASP..131a8002B} {131, 018002}

\bibitem[\protect\citeauthoryear{{Bohlin}, {M{\'e}sz{\'a}ros}, {Fleming},
  {Gordon}, {Koekemoer}  \& {Kov{\'a}cs}}{{Bohlin} et~al.}{2017}]{bohlin_2017}
{Bohlin} R.~C.,  {M{\'e}sz{\'a}ros} S.,  {Fleming} S.~W.,  {Gordon} K.~D.,
  {Koekemoer} A.~M.,   {Kov{\'a}cs} J.,  2017, \mn@doi [\aj]
  {10.3847/1538-3881/aa6ba9}, \href
  {https://ui.adsabs.harvard.edu/abs/2017AJ....153..234B} {153, 234}

\bibitem[\protect\citeauthoryear{{Brown} \& {Bethe}}{{Brown} \&
  {Bethe}}{1994}]{Brown1994}
{Brown} G.~E.,  {Bethe} H.~A.,  1994, \mn@doi [\apj] {10.1086/173844}, \href
  {https://ui.adsabs.harvard.edu/abs/1994ApJ...423..659B} {423, 659}

\bibitem[\protect\citeauthoryear{{Chakrabarti} et~al.,}{{Chakrabarti}
  et~al.}{2023}]{chakrabarti_2023}
{Chakrabarti} S.,  et~al., 2023, \mn@doi [The Astronomical Journal]
  {10.3847/1538-3881/accf21}, \href
  {https://ui.adsabs.harvard.edu/abs/2023AJ....166....6C} {166, 6}

\bibitem[\protect\citeauthoryear{{Chambers} et~al.,}{{Chambers}
  et~al.}{2016}]{chambers_2016}
{Chambers} K.~C.,  et~al., 2016, \mn@doi [arXiv e-prints]
  {10.48550/arXiv.1612.05560}, \href
  {https://ui.adsabs.harvard.edu/abs/2016arXiv161205560C} {p. arXiv:1612.05560}

\bibitem[\protect\citeauthoryear{{Chawla}, {Chatterjee}, {Breivik}, {Moorthy},
  {Andrews}  \& {Sanderson}}{{Chawla} et~al.}{2022}]{Chawla2022}
{Chawla} C.,  {Chatterjee} S.,  {Breivik} K.,  {Moorthy} C.~K.,  {Andrews}
  J.~J.,   {Sanderson} R.~E.,  2022, \mn@doi [\apj] {10.3847/1538-4357/ac60a5},
  \href {https://ui.adsabs.harvard.edu/abs/2022ApJ...931..107C} {931, 107}

\bibitem[\protect\citeauthoryear{{Chubak}, {Marcy}, {Fischer}, {Howard},
  {Isaacson}, {Johnson}  \& {Wright}}{{Chubak} et~al.}{2012}]{Chubak2012}
{Chubak} C.,  {Marcy} G.,  {Fischer} D.~A.,  {Howard} A.~W.,  {Isaacson} H.,
  {Johnson} J.~A.,   {Wright} J.~T.,  2012, \mn@doi [arXiv e-prints]
  {10.48550/arXiv.1207.6212}, \href
  {https://ui.adsabs.harvard.edu/abs/2012arXiv1207.6212C} {p. arXiv:1207.6212}

\bibitem[\protect\citeauthoryear{{Clavel}, {Dubus}, {Casares}  \&
  {Babusiaux}}{{Clavel} et~al.}{2021}]{Clavel2021}
{Clavel} M.,  {Dubus} G.,  {Casares} J.,   {Babusiaux} C.,  2021, \mn@doi
  [\aap] {10.1051/0004-6361/202039317}, \href
  {https://ui.adsabs.harvard.edu/abs/2021A&A...645A..72C} {645, A72}

\bibitem[\protect\citeauthoryear{{Conroy} et~al.,}{{Conroy}
  et~al.}{2020}]{conroy_2020}
{Conroy} K.~E.,  et~al., 2020, \mn@doi [\apjs] {10.3847/1538-4365/abb4e2},
  \href {https://ui.adsabs.harvard.edu/abs/2020ApJS..250...34C} {250, 34}

\bibitem[\protect\citeauthoryear{{Corral-Santana}, {Casares},
  {Mu{\~n}oz-Darias}, {Bauer}, {Mart{\'\i}nez-Pais}  \&
  {Russell}}{{Corral-Santana} et~al.}{2016}]{Corral-Santana2016}
{Corral-Santana} J.~M.,  {Casares} J.,  {Mu{\~n}oz-Darias} T.,  {Bauer} F.~E.,
  {Mart{\'\i}nez-Pais} I.~G.,   {Russell} D.~M.,  2016, \mn@doi [\aap]
  {10.1051/0004-6361/201527130}, \href
  {https://ui.adsabs.harvard.edu/abs/2016A&A...587A..61C} {587, A61}

\bibitem[\protect\citeauthoryear{{Czesla}, {Schr{\"o}ter}, {Schneider},
  {Huber}, {Pfeifer}, {Andreasen}  \& {Zechmeister}}{{Czesla}
  et~al.}{2019}]{pya}
{Czesla} S.,  {Schr{\"o}ter} S.,  {Schneider} C.~P.,  {Huber} K.~F.,  {Pfeifer}
  F.,  {Andreasen} D.~T.,   {Zechmeister} M.,  2019, {PyA: Python
  astronomy-related packages} (\mn@eprint {ascl} {1906.010})

\bibitem[\protect\citeauthoryear{{Eaton} \& {Hall}}{{Eaton} \&
  {Hall}}{1979}]{Eaton1979}
{Eaton} J.~A.,  {Hall} D.~S.,  1979, \mn@doi [\apj] {10.1086/156800}, \href
  {https://ui.adsabs.harvard.edu/abs/1979ApJ...227..907E} {227, 907}

\bibitem[\protect\citeauthoryear{{Eggleton}}{{Eggleton}}{2012}]{eggleton_2012}
{Eggleton} P.~P.,  2012, \mn@doi [Journal of Astronomy and Space Sciences]
  {10.5140/JASS.2012.29.2.145}, \href
  {https://ui.adsabs.harvard.edu/abs/2012JASS...29..145E} {29, 145}

\bibitem[\protect\citeauthoryear{El-Badry \& Burdge}{El-Badry \&
  Burdge}{2022}]{el_badry_burdge_2022}
El-Badry K.,  Burdge K.~B.,  2022, \mn@doi [Monthly Notices of the Royal
  Astronomical Society: Letters] {10.1093/mnrasl/slab135}, 511, 24

\bibitem[\protect\citeauthoryear{El-Badry \& Quataert}{El-Badry \&
  Quataert}{2021}]{el_badry_2021}
El-Badry K.,  Quataert E.,  2021, \mn@doi [Monthly Notices of the Royal
  Astronomical Society] {10.1093/mnras/stab285}, 502, 3436

\bibitem[\protect\citeauthoryear{El-Badry \& Rix}{El-Badry \&
  Rix}{2022}]{badry_rix_2022}
El-Badry K.,  Rix H.-W.,  2022, \mn@doi [Monthly Notices of the Royal
  Astronomical Society] {10.1093/mnras/stac1797}, 515, 1266

\bibitem[\protect\citeauthoryear{El-Badry, Burdge  \& Mróz}{El-Badry
  et~al.}{2022a}]{el_badry_2022a}
El-Badry K.,  Burdge K.~B.,   Mróz P.,  2022a, \mn@doi [Monthly Notices of the
  Royal Astronomical Society] {10.1093/mnras/stac274}, 511, 3089

\bibitem[\protect\citeauthoryear{{El-Badry}, {Seeburger}, {Jayasinghe}, {Rix},
  {Almada}, {Conroy}, {Price-Whelan}  \& {Burdge}}{{El-Badry}
  et~al.}{2022b}]{el_badry_2022b}
{El-Badry} K.,  {Seeburger} R.,  {Jayasinghe} T.,  {Rix} H.-W.,  {Almada} S.,
  {Conroy} C.,  {Price-Whelan} A.~M.,   {Burdge} K.,  2022b, \mn@doi [\mnras]
  {10.1093/mnras/stac815}, \href
  {https://ui.adsabs.harvard.edu/abs/2022MNRAS.512.5620E} {512, 5620}

\bibitem[\protect\citeauthoryear{El-Badry, Conroy, Fuller, Kiman, van Roestel,
  Rodriguez  \& Burdge}{El-Badry et~al.}{2022c}]{el_badry_2022c}
El-Badry K.,  Conroy C.,  Fuller J.,  Kiman R.,  van Roestel J.,  Rodriguez
  A.~C.,   Burdge K.~B.,  2022c, \mn@doi [Monthly Notices of the Royal
  Astronomical Society] {10.1093/mnras/stac2945}, 517, 4916

\bibitem[\protect\citeauthoryear{El-Badry et~al.,}{El-Badry
  et~al.}{2022d}]{gaia_bh1}
El-Badry K.,  et~al., 2022d, \mn@doi [Monthly Notices of the Royal Astronomical
  Society] {10.1093/mnras/stac3140}, 518, 1057–1085

\bibitem[\protect\citeauthoryear{{El-Badry} et~al.,}{{El-Badry}
  et~al.}{2023}]{gaia_bh2}
{El-Badry} K.,  et~al., 2023, \mn@doi [Monthly Notices of the Royal
  Astronomical Society] {10.1093/mnras/stad799}, \href
  {https://ui.adsabs.harvard.edu/abs/2023MNRAS.521.4323E} {521, 4323}

\bibitem[\protect\citeauthoryear{Eyer et~al.,}{Eyer et~al.}{2022}]{eyer_2022}
Eyer L.,  et~al., 2022, Gaia Data Release 3. Summary of the variability
  processing and analysis, \mn@doi{10.48550/ARXIV.2206.06416}, \url
  {https://arxiv.org/abs/2206.06416}

\bibitem[\protect\citeauthoryear{Foreman-Mackey, Hogg, Lang  \&
  Goodman}{Foreman-Mackey et~al.}{2013}]{Foreman-Mackey_2013}
Foreman-Mackey D.,  Hogg D.~W.,  Lang D.,   Goodman J.,  2013, \mn@doi
  [Publications of the Astronomical Society of the Pacific] {10.1086/670067},
  125, 306

\bibitem[\protect\citeauthoryear{Fu, Gu, Zhang, Yi, Qi, Zheng  \& Liu}{Fu
  et~al.}{2022}]{fu_2022}
Fu J.-B.,  Gu W.-M.,  Zhang Z.-X.,  Yi T.,  Qi S.-Y.,  Zheng L.-L.,   Liu J.,
  2022, \mn@doi [The Astrophysical Journal] {10.3847/1538-4357/ac9b4c}, 940,
  126

\bibitem[\protect\citeauthoryear{{Gaia Collaboration} et~al.,}{{Gaia
  Collaboration} et~al.}{2022}]{arenou_2022}
{Gaia Collaboration} et~al., 2022, \mn@doi [arXiv e-prints]
  {10.48550/arXiv.2206.05595}, \href
  {https://ui.adsabs.harvard.edu/abs/2022arXiv220605595G} {p. arXiv:2206.05595}

\bibitem[\protect\citeauthoryear{Gomel, Faigler  \& Mazeh}{Gomel
  et~al.}{2020}]{gomel_faigler_mazeh_2020}
Gomel R.,  Faigler S.,   Mazeh T.,  2020, \mn@doi [Monthly Notices of the Royal
  Astronomical Society] {10.1093/mnras/staa3305}, 501, 2822–2832

\bibitem[\protect\citeauthoryear{Gomel, Faigler  \& Mazeh}{Gomel
  et~al.}{2021a}]{gomel_faigler_mazeh_2021}
Gomel R.,  Faigler S.,   Mazeh T.,  2021a, \mn@doi [Monthly Notices of the
  Royal Astronomical Society] {10.1093/mnras/stab1047}, 504, 2115–2121

\bibitem[\protect\citeauthoryear{{Gomel}, {Faigler}, {Mazeh}  \&
  {Pawlak}}{{Gomel} et~al.}{2021b}]{Gomel2021}
{Gomel} R.,  {Faigler} S.,  {Mazeh} T.,   {Pawlak} M.,  2021b, \mn@doi [\mnras]
  {10.1093/mnras/stab1235}, \href
  {https://ui.adsabs.harvard.edu/abs/2021MNRAS.504.5907G} {504, 5907}

\bibitem[\protect\citeauthoryear{Gomel et~al.,}{Gomel
  et~al.}{2022}]{gomel_gaia_2022}
Gomel R.,  et~al., 2022, \mn@doi [Astronomy \& Astrophysics]
  {10.1051/0004-6361/202243626}

\bibitem[\protect\citeauthoryear{Gordon et~al.,}{Gordon
  et~al.}{2022}]{gordon_2022}
Gordon I.,  et~al., 2022, \mn@doi [Journal of Quantitative Spectroscopy and
  Radiative Transfer] {https://doi.org/10.1016/j.jqsrt.2021.107949}, 277,
  107949

\bibitem[\protect\citeauthoryear{{Gray}}{{Gray}}{1992}]{gray_1992}
{Gray} D.~F.,  1992, {The observation and analysis of stellar photospheres.}.
 Cambridge Astrophysics Vol. 20

\bibitem[\protect\citeauthoryear{{Green}, {Schlafly}, {Zucker}, {Speagle}  \&
  {Finkbeiner}}{{Green} et~al.}{2019}]{green_2019}
{Green} G.~M.,  {Schlafly} E.,  {Zucker} C.,  {Speagle} J.~S.,   {Finkbeiner}
  D.,  2019, \mn@doi [\apj] {10.3847/1538-4357/ab5362}, \href
  {https://ui.adsabs.harvard.edu/abs/2019ApJ...887...93G} {887, 93}

\bibitem[\protect\citeauthoryear{Halbwachs et~al.,}{Halbwachs
  et~al.}{2022}]{halbwachs_2022}
Halbwachs J.-L.,  et~al., 2022, Gaia Data Release 3. Astrometric binary star
  processing, \mn@doi{10.48550/ARXIV.2206.05726}, \url
  {https://arxiv.org/abs/2206.05726}

\bibitem[\protect\citeauthoryear{{Janssens} et~al.,}{{Janssens}
  et~al.}{2022}]{janssens_2022}
{Janssens} S.,  et~al., 2022, \mn@doi [\aap] {10.1051/0004-6361/202141866},
  \href {https://ui.adsabs.harvard.edu/abs/2022A&A...658A.129J} {658, A129}

\bibitem[\protect\citeauthoryear{{Kobulnicky}, {Molnar}, {Cook}  \&
  {Henderson}}{{Kobulnicky} et~al.}{2022}]{kobulnicky_2022}
{Kobulnicky} H.~A.,  {Molnar} L.~A.,  {Cook} E.~M.,   {Henderson} L.~E.,  2022,
  \mn@doi [\apjs] {10.3847/1538-4365/ac75bd}, \href
  {https://ui.adsabs.harvard.edu/abs/2022ApJS..262...12K} {262, 12}

\bibitem[\protect\citeauthoryear{{Kollmeier} et~al.,}{{Kollmeier}
  et~al.}{2017}]{kollmeier_2017}
{Kollmeier} J.~A.,  et~al., 2017, \mn@doi [arXiv e-prints]
  {10.48550/arXiv.1711.03234}, \href
  {https://ui.adsabs.harvard.edu/abs/2017arXiv171103234K} {p. arXiv:1711.03234}

\bibitem[\protect\citeauthoryear{Lurie et~al.,}{Lurie
  et~al.}{2017}]{Lurie_2017}
Lurie J.~C.,  et~al., 2017, \mn@doi [The Astronomical Journal]
  {10.3847/1538-3881/aa974d}, 154, 250

\bibitem[\protect\citeauthoryear{{McClintock} \& {Remillard}}{{McClintock} \&
  {Remillard}}{2006}]{McClintock2006}
{McClintock} J.~E.,  {Remillard} R.~A.,  2006, in , Vol.~39, Compact stellar
  X-ray sources.
pp 157--213

\bibitem[\protect\citeauthoryear{Morris \& Naftilan}{Morris \&
  Naftilan}{1993}]{morris_naftilan_1993}
Morris S.~L.,  Naftilan S.~A.,  1993, \mn@doi [The Astrophysical Journal]
  {10.1086/173488}, 419, 344

\bibitem[\protect\citeauthoryear{Mowlavi et~al.,}{Mowlavi
  et~al.}{2022}]{mowlavi_2022}
Mowlavi N.,  et~al., 2022, Gaia Data Release 3. The first Gaia catalogue of
  eclipsing binary candidates, \mn@doi{10.48550/ARXIV.2211.00929}, \url
  {https://arxiv.org/abs/2211.00929}

\bibitem[\protect\citeauthoryear{{O'Connell}}{{O'Connell}}{1951}]{oconnell_1951}
{O'Connell} D.~J.~K.,  1951, Publications of the Riverview College Observatory,
  \href {https://ui.adsabs.harvard.edu/abs/1951PRCO....2...85O} {2, 85}

\bibitem[\protect\citeauthoryear{Oke \& Gunn}{Oke \& Gunn}{1982}]{Oke_1982}
Oke J.~B.,  Gunn J.~E.,  1982, \mn@doi [Publications of the Astronomical
  Society of the Pacific] {10.1086/131027}, 94, 586

\bibitem[\protect\citeauthoryear{Oke et~al.,}{Oke et~al.}{1995}]{Oke_1995}
Oke J.~B.,  et~al., 1995, \mn@doi [Publications of the Astronomical Society of
  the Pacific] {10.1086/133562}, 107, 375

\bibitem[\protect\citeauthoryear{{Pecaut} \& {Mamajek}}{{Pecaut} \&
  {Mamajek}}{2013}]{pecaut_2013}
{Pecaut} M.~J.,  {Mamajek} E.~E.,  2013, \mn@doi [\apjs]
  {10.1088/0067-0049/208/1/9}, \href
  {https://ui.adsabs.harvard.edu/abs/2013ApJS..208....9P} {208, 9}

\bibitem[\protect\citeauthoryear{{Perley}}{{Perley}}{2019}]{perley_2019}
{Perley} D.~A.,  2019, \mn@doi [\pasp] {10.1088/1538-3873/ab215d}, \href
  {https://ui.adsabs.harvard.edu/abs/2019PASP..131h4503P} {131, 084503}

\bibitem[\protect\citeauthoryear{{Pe{\v{s}}ta} \& {Pejcha}}{{Pe{\v{s}}ta} \&
  {Pejcha}}{2022}]{pesta_2022}
{Pe{\v{s}}ta} M.,  {Pejcha} O.,  2022, \mn@doi [arXiv e-prints]
  {10.48550/arXiv.2212.02553}, \href
  {https://ui.adsabs.harvard.edu/abs/2022arXiv221202553P} {p. arXiv:2212.02553}

\bibitem[\protect\citeauthoryear{{Portegies Zwart}, {Verbunt}  \&
  {Ergma}}{{Portegies Zwart} et~al.}{1997}]{PortegiesZwart1997}
{Portegies Zwart} S.~F.,  {Verbunt} F.,   {Ergma} E.,  1997, \aap, \href
  {https://ui.adsabs.harvard.edu/abs/1997A&A...321..207P} {321, 207}

\bibitem[\protect\citeauthoryear{{Pribulla}, {Va{\v{n}}ko}, {Chochol},
  {Hamb{\'a}lek}  \& {Parimucha}}{{Pribulla} et~al.}{2011}]{pribulla_2011}
{Pribulla} T.,  {Va{\v{n}}ko} M.,  {Chochol} D.,  {Hamb{\'a}lek} {\v{L}}.,
  {Parimucha} {\v{S}}.,  2011, \mn@doi [Astronomische Nachrichten]
  {10.1002/asna.201111569}, \href
  {https://ui.adsabs.harvard.edu/abs/2011AN....332..607P} {332, 607}

\bibitem[\protect\citeauthoryear{{Prochaska} et~al.,}{{Prochaska}
  et~al.}{2020}]{Prochaska_2020}
{Prochaska} J.,  et~al., 2020, \mn@doi [The Journal of Open Source Software]
  {10.21105/joss.02308}, \href
  {https://ui.adsabs.harvard.edu/abs/2020JOSS....5.2308P} {5, 2308}

\bibitem[\protect\citeauthoryear{Prusti et~al.,}{Prusti
  et~al.}{2016}]{prusti_2016}
Prusti T.,  et~al., 2016, \mn@doi [Astronomy \& Astrophysics]
  {10.1051/0004-6361/201629272}, 595

\bibitem[\protect\citeauthoryear{{Pr{\v{s}}a} \& {Zwitter}}{{Pr{\v{s}}a} \&
  {Zwitter}}{2005}]{Prvsa_Zwitter_2005}
{Pr{\v{s}}a} A.,  {Zwitter} T.,  2005, \mn@doi [\apj] {10.1086/430591}, \href
  {https://ui.adsabs.harvard.edu/abs/2005ApJ...628..426P} {628, 426}

\bibitem[\protect\citeauthoryear{{Rigault}}{{Rigault}}{2018}]{rigault_ztfquery}
{Rigault} M.,  2018, {ztfquery, a python tool to access ZTF data}, Zenodo,
  \mn@doi{10.5281/zenodo.1345222}

\bibitem[\protect\citeauthoryear{{Shahaf}, {Bashi}, {Mazeh}, {Faigler},
  {Arenou}, {El-Badry}  \& {Rix}}{{Shahaf} et~al.}{2023}]{shahaf_2023}
{Shahaf} S.,  {Bashi} D.,  {Mazeh} T.,  {Faigler} S.,  {Arenou} F.,  {El-Badry}
  K.,   {Rix} H.~W.,  2023, \mn@doi [\mnras] {10.1093/mnras/stac3290}, \href
  {https://ui.adsabs.harvard.edu/abs/2023MNRAS.518.2991S} {518, 2991}

\bibitem[\protect\citeauthoryear{Sheinis, Bolte, Epps, Kibrick, Miller,
  Radovan, Bigelow  \& Sutin}{Sheinis et~al.}{2002}]{Sheinis_2002}
Sheinis A.~I.,  Bolte M.,  Epps H.~W.,  Kibrick R.~I.,  Miller J.~S.,  Radovan
  M.~V.,  Bigelow B.~C.,   Sutin B.~M.,  2002, \mn@doi [Publications of the
  Astronomical Society of the Pacific] {10.1086/341706}, 114, 851

\bibitem[\protect\citeauthoryear{Vallenari, Brown  \& Prusti}{Vallenari
  et~al.}{2022}]{vallenari_2022}
Vallenari A.,  Brown A.,   Prusti T.,  2022, \mn@doi [Astronomy \&
  Astrophysics] {10.1051/0004-6361/202243940}

\bibitem[\protect\citeauthoryear{{Warner}}{{Warner}}{2003}]{Warner2003}
{Warner} B.,  2003, {Cataclysmic Variable Stars},
  \mn@doi{10.1017/CBO9780511586491.
}

\bibitem[\protect\citeauthoryear{{Wilson} \& {Devinney}}{{Wilson} \&
  {Devinney}}{1971}]{Wilson_Devinney_1971}
{Wilson} R.~E.,  {Devinney} E.~J.,  1971, \mn@doi [\apj] {10.1086/150986},
  \href {https://ui.adsabs.harvard.edu/abs/1971ApJ...166..605W} {166, 605}

\bibitem[\protect\citeauthoryear{{Zahn}}{{Zahn}}{1977}]{Zahn1977}
{Zahn} J.~P.,  1977, \aap, \href
  {https://ui.adsabs.harvard.edu/abs/1977A&A....57..383Z} {57, 383}

\makeatother
\end{thebibliography}

%\begin{thebibliography}{99}
%\end{thebibliography}

%%%%%%%%%%%%%%%%%%%%%%%%%%%%%%%%%%%%%%%%%%%%%%%%%%

%%%%%%%%%%%%%%%%% APPENDICES %%%%%%%%%%%%%%%%%%%%%

\appendix

\section{Rejection of Candidates}
\label{appendix:rejection}

In Table \ref{tab:rejection}, we present a summary of our reasons for manually rejecting promising ellipsoidal variables (i.e. $\hat{q}_{\min} > 0.8$) that survived our initial automated vetting process ($\delta > -30^{\circ}$ and relative parallax / error $>$ 3). The reasons are generally due to the variable being overluminous compared to the main sequence on the CMD, or visual inspection finding irregularities in the ZTF light curves. 

\begin{table*}
\caption{Summary of reasons for rejection of promising candidates ($\hat{q}_{\min} > 0.8$) in vetted sample (apparent $G$-band magnitude $< 17$, \texttt{parallax\_over\_error} $> 3$, and well-constrained extinctions from the 3D dust map of \citet{green_2019}).}
\label{tab:rejection}
\begin{tabular}{cc}
\hline
\textit{Gaia} Source ID & Reason for Rejection \\
\hline
1842474999287779456 & Above main sequence on CMD \\
246694542008676224 & Large difference between light curve minima \\
6858874244124457728 & Not enough ZTF data \\
4115129387457079808 & Above main sequence on CMD \\
4090783421259714048 & Sharp variation in $r$-band ZTF light curve \\
4056769479221855872 & Not enough ZTF data \\
2180310877623317632 & Above main sequence on CMD \\
4513065361686161408 & Poorly sampled ZTF light curves \\
4069146883471008384 & Above main sequence on CMD \\
4245202918693685888 & Not enough ZTF data \\
2118894735067533568 & Above main sequence on CMD \\
4294656413422555136 & Evolved onto giant branch \\
3415519106594116864 & Large difference between light curve minima \\
4268159896853550720 & Not enough ZTF data \\
2019379999263502080 & Above main sequence on CMD \\
4308281733324882688 & Noisy $r$-band ZTF light curve \\
4316223093502225920 & Noisy ZTF light curves \\
4308186591192260224 & Above main sequence on CMD \\
4062968796360712192 & Not enough ZTF data \\
6033056688573451904 & Noisy $g$-band ZTF light curve \\
4063595075664390656 & Noisy ZTF light curves \\
4044200716984124800 & Not enough ZTF data \\
4602733180577371136 & Above main sequence on CMD \\
4505025384714693632 & Noisy ZTF light curves \\
2065168057153338752 & Atypical $r$-band ZTF light curve \\
4320737104132599040 & Above main sequence on CMD \\
237183422828574208 & Large difference between light curve minima \\
187446224071740544 & Above main sequence on CMD \\
4302060043674728320 & Above main sequence on CMD \\
3309994169609595264 & Above main sequence on CMD \\
439282146866334848 & Above main sequence on CMD \\
4096747756468453888 & Above main sequence on CMD \\
4067422028923930880 & Evolved onto giant branch \\
3019860195232390528 & Noisy ZTF light curves \\
3389233807958342912 & Poorly sampled ZTF light curves \\
4310887438426691840 & Significant difference between light curve maxima \\
448105693321340800 & Above main sequence on CMD \\
4107013003158530048 & Above main sequence on CMD \\
504992569522209792 & Above main sequence on CMD \\
3427530553869533440 & Atypical $r$-band ZTF light curve \\
2179940956377621888 & Not enough ZTF data \\
4117615245815921024 & Noisy $g$-band ZTF light curve \\
4323490762226799232 & Not enough ZTF data \\
4151452613074058880 & Not enough ZTF data \\
4124801898687751040 & Above main sequence on CMD \\
4068633617699998080 & Not enough ZTF data \\
1731244348571923200 & Above main sequence on CMD \\
4067808571662706688 & Not enough ZTF data \\
4143833001843654912 & Above main sequence on CMD \\
4070409432055253760 & Not enough ZTF data \\
4056017172771375616 & Not enough ZTF data \\
6038453229081523072 & Not enough ZTF data \\
\hline
\end{tabular}
\end{table*}

\section{Target Selection}
\label{appendix:selection}

In Table \ref{tab:sample}, we provide the coordinates, brightness, and modified minimum mass ratio of all target ellipsoidal variables in our vetted sample ($N = 14$).

\begin{table*}
\caption{Final target list of ellipsoidal binaries with potential compact object companions.}
\label{tab:sample}
\begin{tabular}{ccccccc}
\hline
\textit{Gaia} Source ID & RA (hms) & Dec (dms) & Apparent G Magnitude & Period & $\hat{q}_{\min}$ & Peak-to-Peak Amplitude \\
\hline
2033460169042378624 & 19 51 11.59 & +30 40 22.10 & 16.97 & $0.270854\pm0.000057$ & $0.91\pm0.22$ & 0.169\\[4pt]
1342190925113500416 & 17 38 36.73 & +36 49 21.43 & 16.82 & $0.279545\pm0.000078$ & $0.52\pm0.12$ & 0.137\\[4pt]
3396537760623971456 & 05 46 08.40 & +17 15 39.10 & 16.49 & $0.311336\pm0.000074$ & $0.69\pm0.17$ & 0.180\\[4pt]
4277589209792837632 & 18 14 39.86 & +02 06 01.49 & 16.12 & $0.329189\pm0.000082$ & $1.18\pm0.34$ & 0.165\\[4pt]
2057298203894617216 & 20 22 57.10 & +36 16 52.38 & 16.76 & $0.352784\pm0.000096$ & $1.01\pm0.26$ & 0.177\\[4pt]
2211471926201876992 & 22 54 54.66 & +65 02 53.18 & 16.81 & $0.353515\pm0.000108$ & $0.82\pm0.20$ & 0.163\\[4pt]
4316645409024264576 & 19 29 23.02 & +13 53 40.43 & 16.99 & $0.390803\pm0.000094$ & $1.01\pm0.25$ & 0.135\\[4pt]
188218592331364736  & 05 14 08.92 & +38 49 16.56 & 16.58 & $0.395750\pm0.000142$ & $1.08\pm0.27$ & 0.177\\[4pt]
4305376170746099584 & 19 50 46.50 & +13 42 34.64 & 16.97 & $0.409171\pm0.000104$ & $1.07\pm0.29$ & 0.169\\[4pt]
3456522854428709632 & 06 00 35.88 & +35 58 00.84 & 16.68 & $0.430926\pm0.000134$ & $1.04\pm0.27$ & 0.158\\[4pt]
3457961423656077056 & 05 52 35.93 & +38 31 13.35 & 16.48  & $0.458269\pm0.000152$ & $1.14\pm0.28$ & 0.202\\[4pt]
2163873350591724032 & 20 59 40.91 & +47 16 10.99 & 16.97 & $0.481034\pm0.000174$ & $1.04\pm0.26$ & 0.198\\[4pt]
2062434366319612672 & 20 13 10.48 & +40 24 56.28 & 16.36 & $0.721862\pm0.000404$ & $1.06\pm0.26$ & 0.193\\[4pt]
4310698008860121216 & 19 06 17.45 & +09 34 08.08 & 16.99 & $0.751384\pm0.000436$ & $1.62\pm0.46$ & 0.170 \\
\hline
\end{tabular}
\end{table*}

\section{Light Curves}
\label{appendix:lcs}

\begin{figure*}
\includegraphics[width=0.66\textwidth]{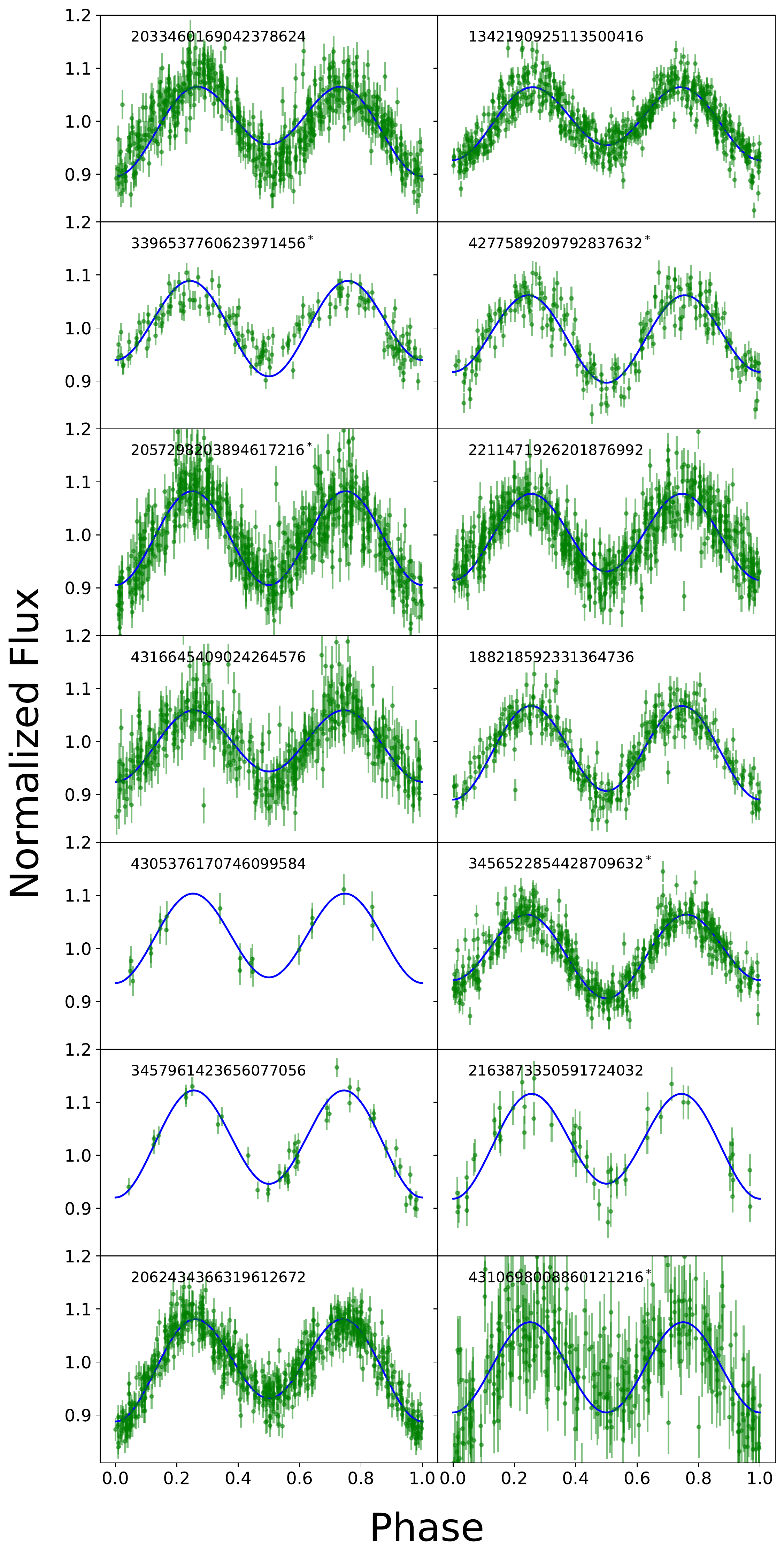}
\caption{Comparison of $g$-band ZTF light curves for all observed ellipsoidal variables.}
\label{fig:all_lcs}
\end{figure*}

We provide all of the ZTF $g$-band light curves for reference in Figure \ref{fig:all_lcs}. We assume that the a light curve displaying ellipsoidal variability can be described using a sum of harmonics:

\begin{equation}
    g = \bar{g} + \sum_{i = 1}^2 a_{i} \cos{\left(\frac{2 \pi i (t - T_0)}{P}\right)}
\end{equation}

We use non-linear fitting techniques (i.e.\ Levenberg-Marquardt algorithm) to find the mean magnitude $\bar{g}$, best-fit coefficients $a_{i}$, and epoch of conjunction $T_0$ given the \textit{Gaia} $G$-band period $P$, and overplot these fits in Figure \ref{fig:all_lcs}. Candidates for which $T_0$ was later shifted by half a period based on radial velocity measurements are marked by an asterisk.

\section{Radial Velocity Plots}
\label{appendix:rvs}

\begin{figure*}
\includegraphics[width=0.66\textwidth]{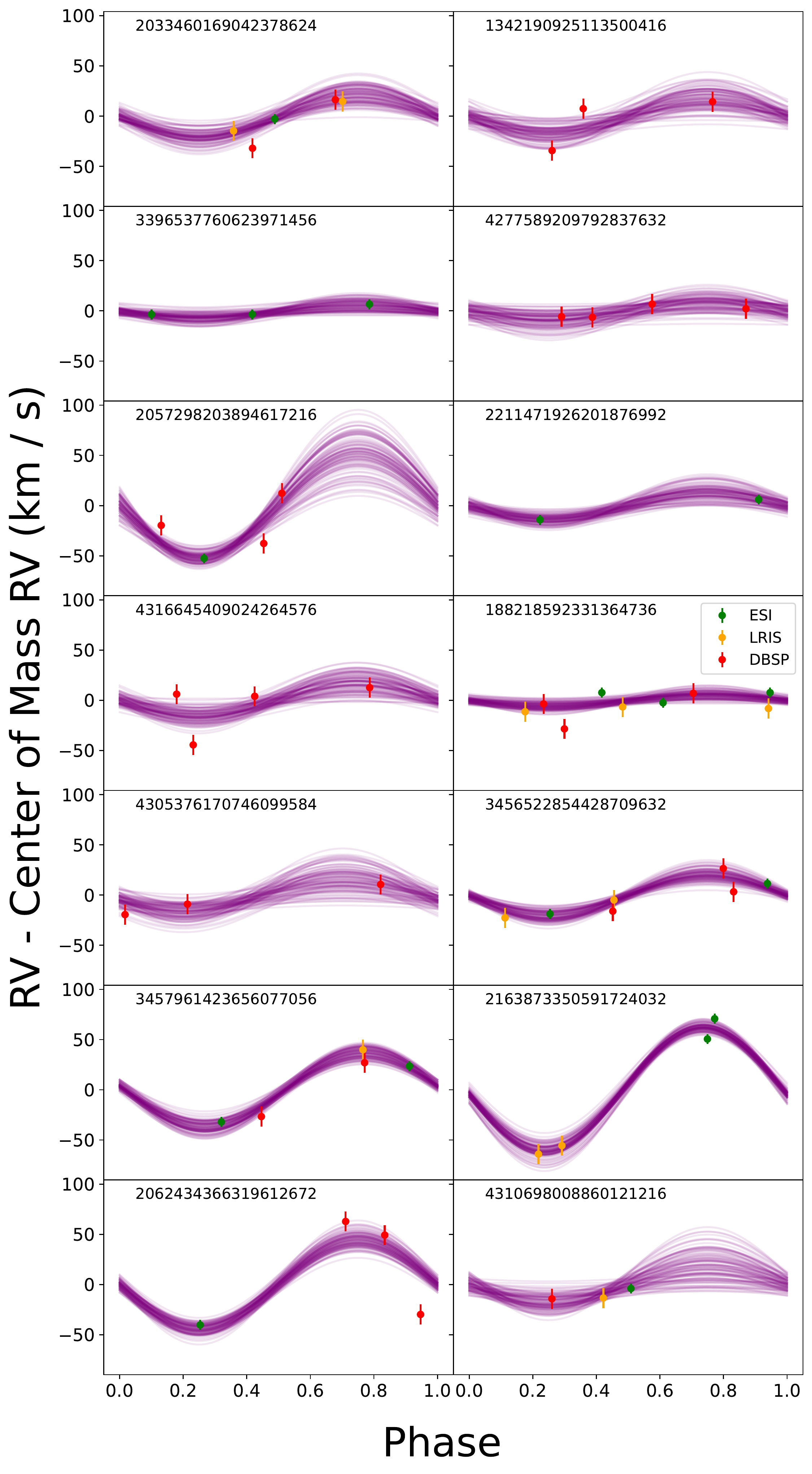}
\caption{All radial velocity curves for observed ellipsoidal variables. Center of mass RVs are subtracted for ease of comparison.}
\label{fig:all_rvs}
\end{figure*}

We provide all of our radial velocity fits for ease of comparison in Figure \ref{fig:all_rvs}. The mean radial velocities are subtracted so that all plots share the same vertical axis. As described in Section \ref{sec:analysis}, color coding is by instrument used for each observation.

%%%%%%%%%%%%%%%%%%%%%%%%%%%%%%%%%%%%%%%%%%%%%%%%%%

% Don't change these lines
\bsp	% typesetting comment
\label{lastpage}
\end{document}